\documentclass[twocolumn]{aastex631}
\usepackage{placeins}
\usepackage{booktabs}
\usepackage{amsmath}

\newcommand\z{\textit{z}}
\newcommand\HST{\textit{HST}}

\shorttitle{Galaxy Mergers in TNG50}
\shortauthors{Schechter et al.}

\graphicspath{{./}{figures/}}

\begin{document}

\title{Enhanced Star Formation and Black Hole Accretion Rates in Galaxy Mergers in IllustrisTNG50}

\correspondingauthor{Aimee Schechter}
\email{aimee.schechter@colorado.edu}

\author[0000-0001-7120-2433]{Aimee L. Schechter}
\affiliation{Department of Astrophysical and Planetary Sciences, 
University of Colorado, 
Boulder, CO 80309, USA}

\author{Shy Genel}
\affiliation{Center for Computational Astrophysics,
Flatiron Institute,
New York, NY 10010, USA}
\affiliation{Columbia Astrophysics Laboratory, Columbia University, 550 West 120th Street, New York, NY, 10027, USA}

\author{Bryan Terrazas}
\affiliation{Department of Physics \& Astronomy, 
Oberlin College, 
Oberlin, OH, 44074, USA}

\author{Julia M. Comerford}
\affiliation{Department of Astrophysical and Planetary Sciences, 
University of Colorado, 
Boulder, CO 80309, USA}

\author[0000-0002-5891-1603]{Abigail Hartley}
\affiliation{Department of Astrophysical and Planetary Sciences, 
University of Colorado, 
Boulder, CO 80309, USA}
\affiliation{Department of Physics, 
Stanford University, 
382 Via Pueblo Mall, Stanford, CA 94305, USA}

\author[0000-0002-6748-6821]{Rachel S. Somerville}
\affiliation{Center for Computational Astrophysics,
Flatiron Institute,
New York, NY 10010, USA}

\author{Rebecca Nevin}
\affiliation{Fermi National Accelerator Laboratory,  
P.O. Box 500, 
Batavia, IL 60510, USA}

\author[0000-0003-1407-6607]{Joseph Simon} 
\altaffiliation{NSF Astronomy \& Astrophysics Postdoctoral Fellow}
\affiliation{Department of Astrophysical and Planetary Sciences, 
University of Colorado, 
Boulder, CO 80309, USA}

\author{Erica Nelson}
\affiliation{Department of Astrophysical and Planetary Sciences, 
University of Colorado, 
Boulder, CO 80309, USA}


\begin{abstract}

\textnormal{Many theoretical and observational studies have suggested that galaxy mergers may trigger enhanced star formation or active galactic nuclei (AGN) activity}.
We present an analysis of merging and nonmerging galaxies from $0.2 \leq z \leq 3$ in the IllustrisTNG50 simulation. 
These galaxies encompass a range of masses ($M_\star > 10^{8}M_\odot$), multiple merger stages, and mass ratios ($\geq1:10$). 
We examine the effect that galaxy mergers have on star formation and black hole accretion rates in the TNG50 universe. 
We additionally investigate how galaxy and black hole mass, merger stage, merger mass ratio, and redshift affect these quantities. 
Mergers in our sample show excess specific star formation rates (sSFR) at $z \leq 3$ and enhanced specific black hole accretion rates (sBHAR) at $z \lesssim 2$. 
The difference between sSFRs and sBHARs in the merging sample compared to the non-merging sample increases as redshift decreases. 
Additionally, we show that these enhancements persist for at least $\sim1$ Gyr after the merger event.
Investigating how mergers behave in the TNG50 \textnormal{simulation} throughout cosmic time enables both a better appreciation of the importance of spatial resolution in cosmological simulations and a better basis to understand our high-$z$ universe with observations from \textit{JWST}.

\end{abstract}

\section{Introduction} \label{sec:intro}
Galaxy mergers are a key component of our picture of galaxy evolution, as galaxies form hierarchically in the $\Lambda$ Cold Dark Matter ($\Lambda$CDM) paradigm. 
Both observational and theoretical work has found that mergers are important for driving numerous evolutionary processes, including star formation (SF) and black hole (BH) accretion. 
During a galaxy merger, simulations show tidal torques drive cold gas into the center of the system, and gas gets compressed to higher densities leading to bursts of star formation, and potentially activating active galactic nuclei (AGN) (e.g., \citealt{barnes_fueling_1991, mihos_gasdynamics_1996, di_matteo_energy_2005, springel_black_2005, springel_formation_2005, springel_modelling_2005}). 
However, observations have revealed a more complicated picture where multiple characteristics of the merger influence its effect on star formation rate (SFR) and black hole accretion rate (BHAR) (e.g., \citealt{ellison_galaxy_2008, patton_galaxy_2011, barrows_census_2023}).

Merger stage, stellar mass ratio, and redshift are three of the key aspects that can affect the amount of star formation or AGN actvity triggered by the galaxy merger.
As we head into the era of high redshift galaxy evolution with the James Webb Space Telescope (\textit{JWST}), it is essential to examine how mergers affect star formation and black hole accretion in galaxies across cosmic time, especially for the smaller, lower mass galaxies that are common in the early universe.

The effects of mergers on star formation and AGN can be studied by simulating isolated galaxy mergers (e.g., \citealt{di_matteo_energy_2005, springel_formation_2005, moreno_interacting_2019, he_molecular_2023}), zoom-in simulations (e.g., \citealt{lovell_first_2021, vijayan_first_2021, sharma_connection_2023}) and cosmological hydrodynamic simulations such as IllustrisTNG50 \citep{nelson_first_2019, pillepich_first_2019}, EAGLE \citep{crain_eagle_2015, schaye_eagle_2015}, Horizon-AGN \citep{dubois_dancing_2014, dubois_horizon-agn_2016, kaviraj_horizon-agn_2017, laigle_horizon-agn_2019}, and SIMBA \citep{dave_simba_2019}).
Zoom-in and isolated merger simulations have the benefit of being able to simulate more realistic interstellar media and resolve smaller star forming clumps.
However, these simulations lack both the cosmological volumes necessary to gather meaningful statistics and the large-scale environments in which galaxies live in cosmological simulations. 

Isolated galaxy merger simulations show that the star formation rate peaks during the first pass, or early stages of the merger \citep{hopkins_cosmological_2008, kim_galaxy_2009, renaud_starbursts_2014} with black hole accretion rates peaking later or closer to coalescence \citep{springel_black_2005, hopkins_dynamical_2012}.
\citet{sharma_connection_2023} \textnormal{find that}
merging host galaxies are not preferentially found in a sample of AGN in zoom-in simulations, except at low-$z$. 
Cosmological simulations such as SIMBA and TNG show both enhanced star formation and black hole activity due to mergers (e.g., \citealt{rodriguez_montero_mergers_2019, patton_interacting_2020, hani_interacting_2020, byrne-mamahit_interacting_2022, byrne-mamahit_interacting_2024}).

Observations remain split on whether major mergers (stellar mass ratio $\geq 1:4$) do trigger AGN (e.g., \citealt{ellison_galaxy_2011, treister_major_2012, satyapal_galaxy_2014, glikman_major_2015, comerford_excess_2024}), or are no more responsible for AGN triggering than other secular processes (e.g., \citealt{georgakakis_host_2009, cisternas_bulk_2011, kocevski_candels_2012, simmons_moderate-luminosity_2012, villforth_morphologies_2014, mechtley_most_2016, villforth_host_2017}).
Meanwhile, minor mergers (stellar mass ratio $<1:4$) may have less of an impact on star formation or AGN accretion than major mergers since there is less disruption to the system as a whole. 
Simulations show that the stellar mass ratio is an important factor in determining how much of an increase in star formation and AGN activity a merger could produce, with major mergers having a larger increase than minor mergers \citep{cox_effect_2008, capelo_growth_2015}.

The connection between mergers, star formation, and AGN may also \textnormal{change with cosmic time}, 
as galaxies generally had higher star formation rates at earlier times (e.g., \citealt{madau_cosmic_2014}), and galaxies had higher gas fractions and more turbulent disks \textnormal{in the past} than they do today \citep{ubler_evolution_2019}. 
Some of the disagreements listed above may be explained by the redshift ranges of observations, but not all. 
A picture including stellar mass, gas fraction, merger stage, and merger mass ratio, is needed to fully understand the connection between mergers, star formation and black hole activity.

High redshift mergers have historically been extremely difficult to identify observationally, but \textit{JWST} enables more high-$z$ merger studies \citep{snyder_automated_2019, rose_identifying_2023}. 
Using multiwavelength data combined with new analysis methods for high redshift merger identification will be crucial to observe all aspects of the merger-AGN connection.
Large cosmological simulations provide a framework for studying the impact of galaxy mergers on star formation and AGN, since galaxies can be studied across cosmic time using the same analysis methods, which is not usually possible with observations.

IllustrisTNG50, with its box of $\sim50$ co-moving Mpc/side and $\sim 0.1$kpc resolution, provides the benefit of following galaxies through cosmic time in realistic environments, while having a spatial resolution that approaches that of zoom-in simulations.
TNG50 provides a unique opportunity to compare both to results from the larger box runs, TNG100 and TNG300 ($\sim 100$ and $\sim 300$ co-moving Mpc per side respectively), and to zoom-in simulations. 

Multiple papers have studied the connections between galaxy mergers, star formation, and black hole activity in the TNG universe.
\cite{patton_interacting_2020} found that galaxies' specific star formation rates (sSFRs) are enhanced in both TNG100 and TNG300 from nearby companion galaxies, with greater enhancements at low redshift.
Additionally, the sSFR enhancement is largest at small galaxy separations, motivating work investigating if this trend persists in TNG50. 
\cite{hani_interacting_2020} see an overall trend in TNG300 that galaxy mergers just after coalescence exhibit enhanced star formation only if they were star forming to begin with. 

Black holes in TNG are also affected by mergers, with \cite{byrne-mamahit_interacting_2022} finding mergers are more likely to experience an AGN phase than isolated galaxies across $0 < z < 1$. 
This study also found that mergers in TNG300 had a slightly higher black hole accretion rate than those in TNG100.
These results raise the question of how TNG50's increased spatial resolution might affect the measured star formation and black hole accretion rates of merging galaxies, and if the mergers look increasingly different from the nonmergers as the spatial resolution increases.

TNG, with its three box sizes and resolutions, enables a study of how the physical model produces (or does not produce) star formation and black hole accretion enhancements in galaxy mergers, and how this compares to the observed universe. This paper aims to address how the increased spatial resolution of TNG50 affects the star formation and black hole accretion rates in galaxy mergers from $0.2 \leq z \leq 3$ in the TNG50 universe. We also address whether this behavior matches observations. The TNG simulation is described in Section~\ref{sec:data}, our analysis methods are shown in Section~\ref{sec:methods}, and our results are presented in Section~\ref{sec:results}. Section~\ref{sec:discussion} compares our results to previous work about galaxy merger driven SFR and BHAR, both theoretical and observational, and discusses the physics of the TNG50 simulation that are especially important in this work.

\section{Data} \label{sec:data}
\subsection{IllustrisTNG Simulation}
The IllustrisTNG simulation is a cosmological, gravo-magnetohydrodynamical simulation run with the moving mesh code arepo \citep{springel_e_2010} in three box sizes: TNG300, TNG100, and TNG50, which are 300 Mpc, 100 Mpc, and 50 Mpc per box side respectively \citep{springel_first_2018, pillepich_first_2018, nelson_first_2018, marinacci_first_2018, naiman_first_2018, nelson_first_2019, pillepich_first_2019}. 
TNG includes gas, stars, and supermassive black holes in addition to dark matter that it evolves self-consistenly through cosmic time ($127 \leq z \leq 0$) using cosmological parameters consistent with \citet{planck_collaboration_planck_2016}.
This paper utilizes the TNG50-1 run which contains 2160$^3$ resolution elements, a baryonic mass resolution of $8.5\times10^4M_\odot$, and a dark matter mass resolution of $4.5\times10^5M_\odot$. 

TNG has been shown to qualitatively match observations of galaxy colors and morphologies, gas properties, and kinematics \citep{nelson_first_2018, pillepich_first_2019}.
However, the TNG galaxies quench faster than observations \textnormal{indicate is the case for real galaxies}, and have a black hole mass threshold for quenching which is not seen in observations \citep{terrazas_relationship_2020}. 
Galaxies in TNG also have a tighter $M_{\mathrm{BH}} - M_\star$ relationship than observations at the same mass ranges \citep{terrazas_relationship_2020}. 
The AGN luminosity function in TNG is in conflict with observations, overproducing low luminosity AGN and diverging more as redshift increases \citep{habouzit_supermassive_2022}.

Importantly for this work, TNG includes gas radiative processes, stellar evolution and chemical enrichment, black hole formation and evolution, and black hole feedback in two modes: a thermal mode and a kinetic mode \citep{weinberger_simulating_2017}. TNG50's spatial resolution enables us to look at sub-kpc scale processes, such as star formation, along with large, galactic scale processes, such as galaxy mergers, in a cosmological environment. 
Additionally, with this spatial resolution TNG50 can marginally resolve the disk scale height, which the other TNG volumes cannot, enabling more accurate modeling of disk instabilities.

\subsubsection{Star Formation Model}
Each stellar particle represents a single-age stellar population described by a \citet{chabrier_galactic_2003} initial mass function. 
These stellar populations age and redistribute metals back into the interstellar medium (ISM). 
The star formation model has a gas consumption timescale, $t_{\mathrm{sfr}}$ = 2.2 Gyr and a neutral hydrogen atom number density, $\rho_{\mathrm{sfr}}$ = 0.13 cm$^{-3}$ \citep{springel_cosmological_2003, vogelsberger_model_2013}.
When the gas exceeds this density, the gas cells are converted to stellar particles.
In order to avoid extremely small time steps at the TNG50 resolution, a change was made to the effective equation of state of star forming gas, but this was shown to have a negligible effect on galaxy properties \citep{nelson_first_2019}.
\subsubsection{Black Hole and Accretion Model}
The full black hole and feedback prescription is given in \cite{weinberger_simulating_2017}, and we provide a summary here. 
\textnormal{TNG includes a thermal feedback mode that is activated when the BH is in a high accretion rate state, and a kinetic feedback mode that occurs when the BH is in low accretion rate state}. 
The accretion rate is determined by the Eddington limited Bondi rate, where the Bondi and Eddington rates are given by:

\begin{equation} \label{eq:bondi}
    \dot{M}_{\mathrm{Bondi}} = \frac{4\pi G^2M_{\mathrm{BH}}^2\rho}{c_\mathrm{s}^3} 
\end{equation}

\begin{equation} \label{eq:edd}
    \dot{M}_{\mathrm{Edd}} = \frac{4\pi GM_{\mathrm{BH}}m_\mathrm{p}}{\epsilon_\mathrm{r} \sigma_\mathrm{T} c}
\end{equation}
where $\rho$ is the gas density near the black hole, $c_s$ is the sound speed near the black hole, $\epsilon_r$ is the radiative accretion efficiency, and $\sigma_T$ is the Thompson cross section. 
\textnormal{The thermal mode is activated when the ratio of BH accretion rate to the Eddington rate is greater than the BH-mass dependent critical rate  $\chi$,} 
\begin{equation}
\begin{split}
    \frac{M_{\mathrm{Bondi}}}{M_{\mathrm{Edd}}} & \geq \chi \\
    & \geq \text{min}\left[\chi_0 \left(\frac{M_{\mathrm{BH}}}{10^8 M_\odot}\right)^\beta, 0.1\right] 
\end{split}
\end{equation}
where $\chi_0$ and $\beta$ are parameters. \textnormal{Correspondingly, the kinetic mode is activated when the accretion rate is less than this critical value. 
The thermal feedback injects thermal feedback isotropically and uniformly into the gas surrounding the black hole. 
This heats the surrounding gas and can drive outflows.
The kinetic mode injects kinetic energy stochastically into the gas around the black hole, changing the velocity and momentum of those particles. 
This induces bulk motion and turbulence into the gas.}

The bolometric AGN luminosity of a black hole can be calculated using a single expression for all black holes regardless of their Eddington ratio:
\begin{equation}
    L_{\mathrm{bol}} = \frac{\epsilon_r}{1 - \epsilon_r} \dot{M}_{\mathrm{BH}}c^2 
    \label{eq:Lbol}
\end{equation}
where $\epsilon_r$ is the radiative efficiency in TNG ($\epsilon_r$ = 0.2), $\dot{M}_{\mathrm{BH}}$ the BHAR, and $c$ is the speed of light. 
See \citet{habouzit_linking_2019, habouzit_supermassive_2022} for discussions of Eddington ratio and black hole luminosity.

\textnormal{A black hole with mass  $1.2\times10^6 \, \, $M$_\odot$ is seeded in any halo that does not yet contain a black hole once it reaches a mass threshold of $M_{\mathrm{FOF}} = 5\times10^{10}$ M$_\odot$.}
Black holes are shifted to the minimum of the gravitational potential at every global integrated time step 
to ensure no halos lose their central black holes and to speed up mergers.

\section{Methods} \label{sec:methods}
\subsection{Selecting merging and non-merging galaxies} \label{selectgalaxies}
\begin{table*}
\centering
\begin{tabular}{ccccccc}
\toprule

Redshift & Nonmergers & Mergers & Pre-Coalescence & Post-Coalescence & Major & Minor \\
\midrule

3.0 & 2346 & 2346 & 734 & 1612 & 944 & 1402 \\
2.0 & 1373 & 1373 & 414 & 959 & 551 & 822 \\
1.5 & 957 & 957 & 286 & 671 & 411 & 546 \\
1.0 & 575 & 575 & 190 & 385 & 236 & 339 \\
0.7 & 432 & 432 & 137 & 295 & 227 & 205 \\
0.5 & 300 & 300 & 101 & 199 & 164 & 136 \\
0.4 & 329 & 329 & 111 & 218 & 181 & 148 \\
0.3 & 288 & 288 & 100 & 188 & 157 & 131 \\
0.2 & 269 & 269 & 95 & 174 & 150 & 119 \\

\bottomrule
\end{tabular}
\label{table:counts}
\caption{Number of galaxies in each redshift bin by category: nonmerger or merger, and mergers broken down by merger stage or stellar mass ratio. See Section~\ref{sec:methods} for a description of how we define each category.}
\end{table*}

We identify merging and non-merging subhalos in TNG50 using the \texttt{SubLink} merger trees from \citet{rodriguez-gomez_merger_2015}, where a subhalo in TNG is defined as a substructure of locally overdense, gravitationally bound groups of particles identified via the Subfind algorithm (\citealt{springel_populating_2001}).  
We restrict our search to subhalos that have a stellar mass greater than 1000 times the baryonic mass resolution of TNG50-1, which is $5.7\times 10^4 \, h^{-1} \, M_{\odot}$, or $10^{7.9} M_{\odot}$, \textnormal{to ensure that the galaxies in our sample are well resolved}. See Table \ref{table:counts} for a tally of the number of galaxies in each merger category at each redshift bin.

Tracing any given subhalo backward in time will find its progenitors, and tracing that subhalo forward in time will find its descendants.  
We use the definition of a merger of two galaxies from \citet{rodriguez-gomez_merger_2015} to identify merging subhalos, where a merger occurs when a subhalo has more than one direct progenitor. 
Subhalos that have more than one direct progenitor are henceforth referred to as `currently merging' subhalos. 
We broaden this definition to include subhalos that merge within 250 Myr of the `currently merging' snapshot in question as in \citet{snyder_automated_2019}; pre-mergers are first progenitor subhalos that will merge in the next 250 Myr, and post-mergers previously merged within the past 250 Myr. 
Using these merger criteria, we build catalogs of mergers centered at $z = \{0.2, 0.3, 0.4, 0.5, 0.7, 1, 1.5, 2, 3\}$, designed to be spaced equally in time ($\Delta t \sim$1 Gyr).

The merger catalogs include the redshift, the time since merger ($t$ - $t_{\mathrm{merg}}$), where $t_{\mathrm{merg}}$ is the moment where two direct progenitors exist, the stellar mass, and the stellar mass ratio of the merger ($\mu_{*}$) at each galaxy's maximum stellar mass as done in \citet{rodriguez-gomez_merger_2015}. These catalogs include both minor and major mergers, where major mergers have a stellar mass ratio $\mu_* \geq 0.25$, and minor mergers have a stellar mass ratio $0.1 < \mu_* < 0.25$ (\citealt{rodriguez-gomez_merger_2015}).

\begin{figure}[t]
    \centering
    \includegraphics[width=0.45\textwidth]{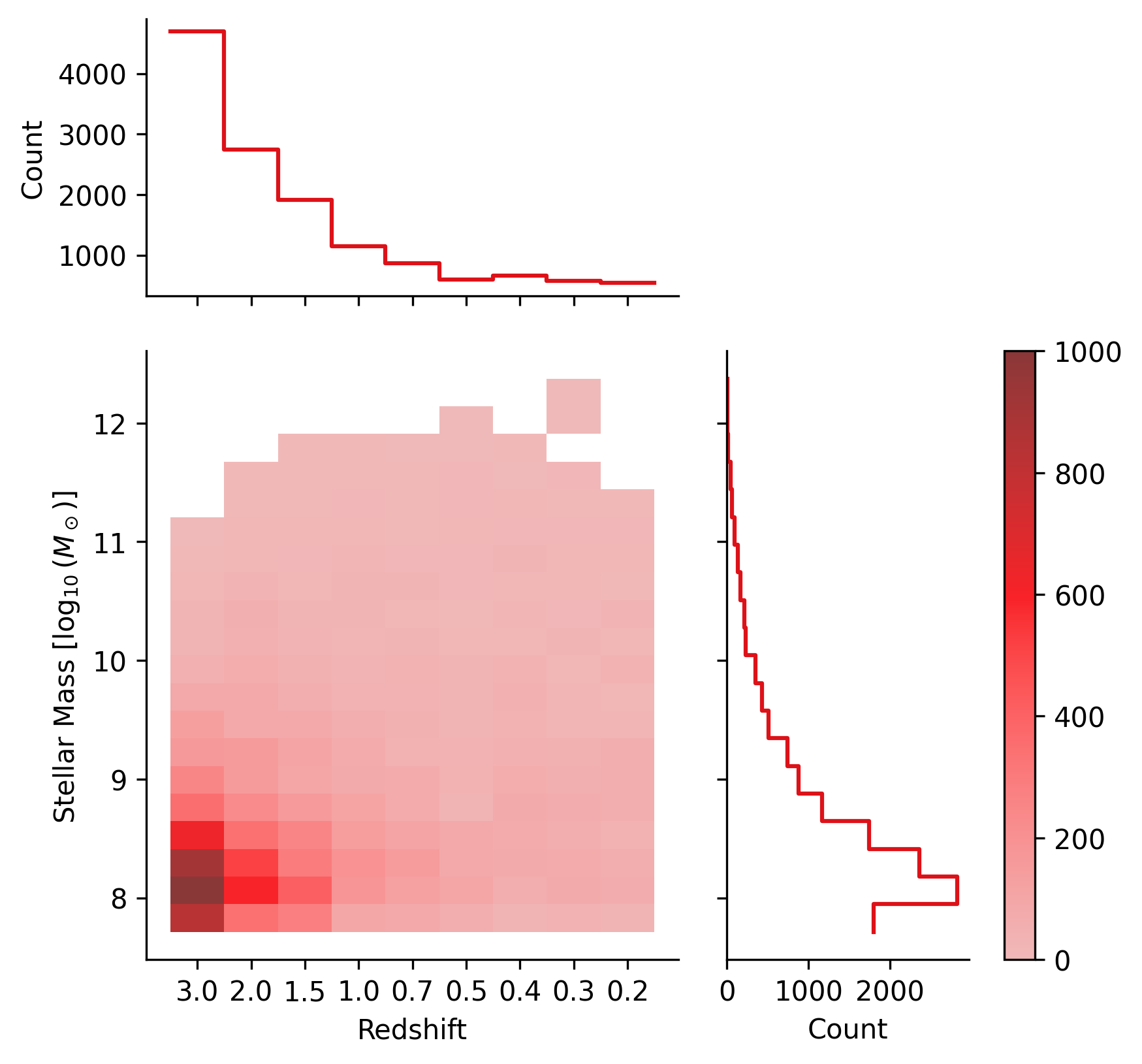}
    \includegraphics[width=0.45\textwidth]{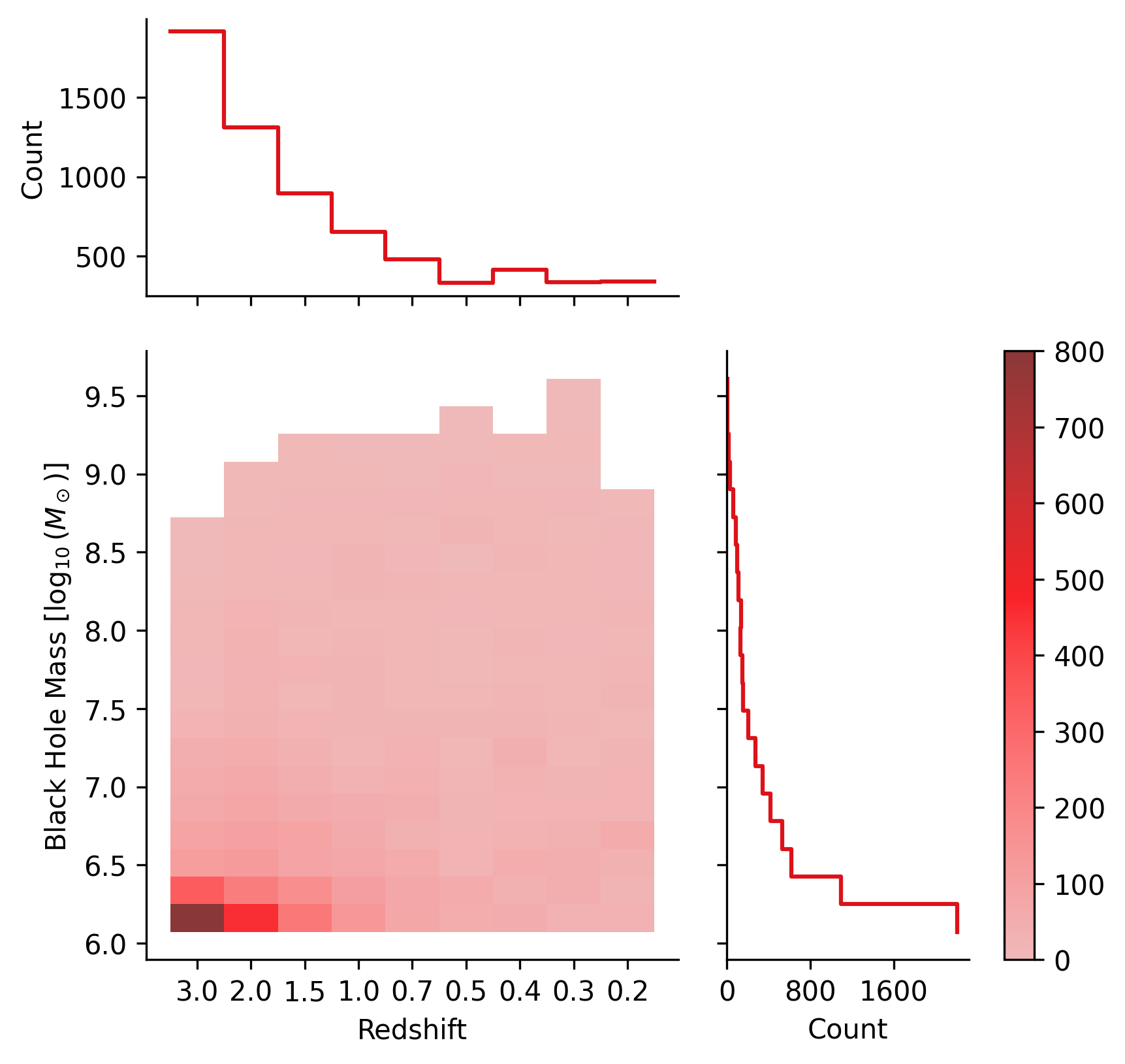}
    \caption{Bivariate histograms showing the number of total galaxies (top) and black holes (bottom) in our sample. The marginal histogram on top shows the total number of galaxies (top) and black holes (bottom) in each redshift bin. The marginal histogram on the right shows the stellar mass distribution of the galaxies (top) and supermassive black hole mass distribution (bottom). Not every galaxy has a seeded black hole, meaning the total number counts are higher in the top plot. There are more galaxies at high redshifts and low masses in our sample, due to the nature of a small box, cosmological simulation, but since the mergers and nonmergers are matched by mass and redshift this will not impact our overall conclusions.}
    \label{fig:Sample}
\end{figure}

To build our matched samples of non-merging galaxies at each redshift bin, each merging galaxy is matched with one non-merging galaxy in logarithmic bins in stellar mass and redshift. 
\textnormal{We use the matching scheme from \citet{bickley_convolutional_2021}, which involves defining a threshold of match parameters in log space and loosening the threshold if no matched galaxies are found. 
The beginning threshold on stellar mass matching is $e^{0.1}$ times the merger's stellar mass.
\textnormal{This gives a tighter constraint than statistical uncertainties on stellar mass measurements \citep{mendel_catalog_2014, patton_galaxy_2016}.}
With each iteration, the tolerance grows by an exponent factor of 1.5.
Once a nonmerger has been selected as a match for a merger, that galaxy is ineligible to be chosen again as a future match to avoid duplicates in the sample.}

After this matching process, any galaxies that were not identified through the merger trees for multiple snapshots in a row were removed from the sample, \textnormal{along with the corresponding matched galaxy}.
 
We obtain the merger tree and group catalogs for each galaxy selected. 
We follow each galaxy along the main progenitor and descendant branch, and get its stellar mass, black hole mass, gas mass, star formation rate, and black hole accretion rate at each snapshot.
All quantities are measured within twice the stellar half mass radius.
Figure \ref{fig:Sample} shows the distributions of stellar and black hole masses in our final sample, and compares the number of galaxies in each redshift and stellar mass bin. 
We note here that since not all subhalos have seeded black holes, the number counts on the black hole sample plot are much lower than that of the galaxy sample plot.

\subsection{Analysis Methods}
We bin the galaxies into two levels of categories.
As a larger umbrella category, the galaxies are binned by mergers and nonmergers for all analysis. 
From here we bin the mergers by merger stage and merger stellar mass ratio. Our merger stage is split into two bins: pre-coalescence mergers and post-coalescence mergers. 
Due to the large snapshot separations in time in a cosmological simulation like TNG50 and the repositioning of the black hole to the center of the potential well, these stages are not completely analogous to those of an observational merger. 
Rather, the pre-coalescence stage is when the SUBFIND algorithm still identifies two separate subhalos, but they will become one within the next snapshot, and the post-coalescence stage is when the SUBFIND algorithm sees one subhalo that has two direct progenitors one or two snapshots previously. 
The galaxies are also binned by stellar and black hole mass in bins of $10^{0.5}M_\odot$ to disentangle the connection of galaxy or black hole mass to star formation and black hole accretion.

For this paper, we are interested in looking at (s)BHARs and (s)SFRs, and how these rates change relative to merger mass ratios or merger stages. 
We separate our galaxy sample into pre-coalescence mergers, post-coalescence mergers, major mergers, minor mergers, and nonmergers. 
Note that a galaxy can be counted in both a merger stage category and a merger mass ratio category, but will never be included twice in the same plot or calculation.
Within each category, we used scipy's binned statistic function to bin the galaxies by stellar or black hole mass (depending on if sSFR or sBHAR is being analyzed) and calculate medians and standard deviations. 

Additionally, we use Hellinger distances to quantify the statistical difference (or lack thereof) in the distributions of parameters in mergers and nonmergers. 
\textnormal{We choose to use a Hellinger distance rather than a KS test due to Lindley's paradox: as the datasets get large, p-values can get arbitrarily small.
Hellinger distances measure the similarity between two distributions, and can be more informative than a KS test for large samples.}

A Hellinger distance is defined as:
\begin{equation}
    H(P, Q) = \frac{1}{\sqrt{2}} \sqrt{\sum_{i = 1}^k \left(\sqrt{p_i} - \sqrt{q_i}\right)^2}
\end{equation}
where $P$ and $Q$ are discrete probability distributions and $p_i$ and $q_i$ are the discrete samples.
\textnormal{It can be thought of as measuring the statistical overlap between two distributions of the same variable.}
A Hellinger distance is always between 0 and 1, with a value of 0 meaning the two distributions are the same, and a value of 1 meaning the two distributions are completely distinct. 

\textnormal{
We calculate a separate Hellinger distance for each of our two variables, sSFR and sBHAR, in each mass and redshift bin. 
\textit{P} and \textit{Q} are the distributions for mergers and nonmergers for that variable, in that bin.
We are then summing over the sSF(BHA)Rs of each individual merging and nonmerging matched pair in the bin, therefore comparing the distributions of a variable (sSFR or sBHAR) between the merging and nonmerging samples. }

\section{Results} \label{sec:results}
We analyze both the sSFRs and sBHARs for all galaxies in our sample, and explore how those of merging galaxies may behave differently to those of isolated, nonmerging galaxies.

\subsection{Excess sSFR and sBHAR in Merging Galaxies}\label{sec:MergersvsNonmergers}

\begin{figure*}
    \centering
    \includegraphics[width=\textwidth]{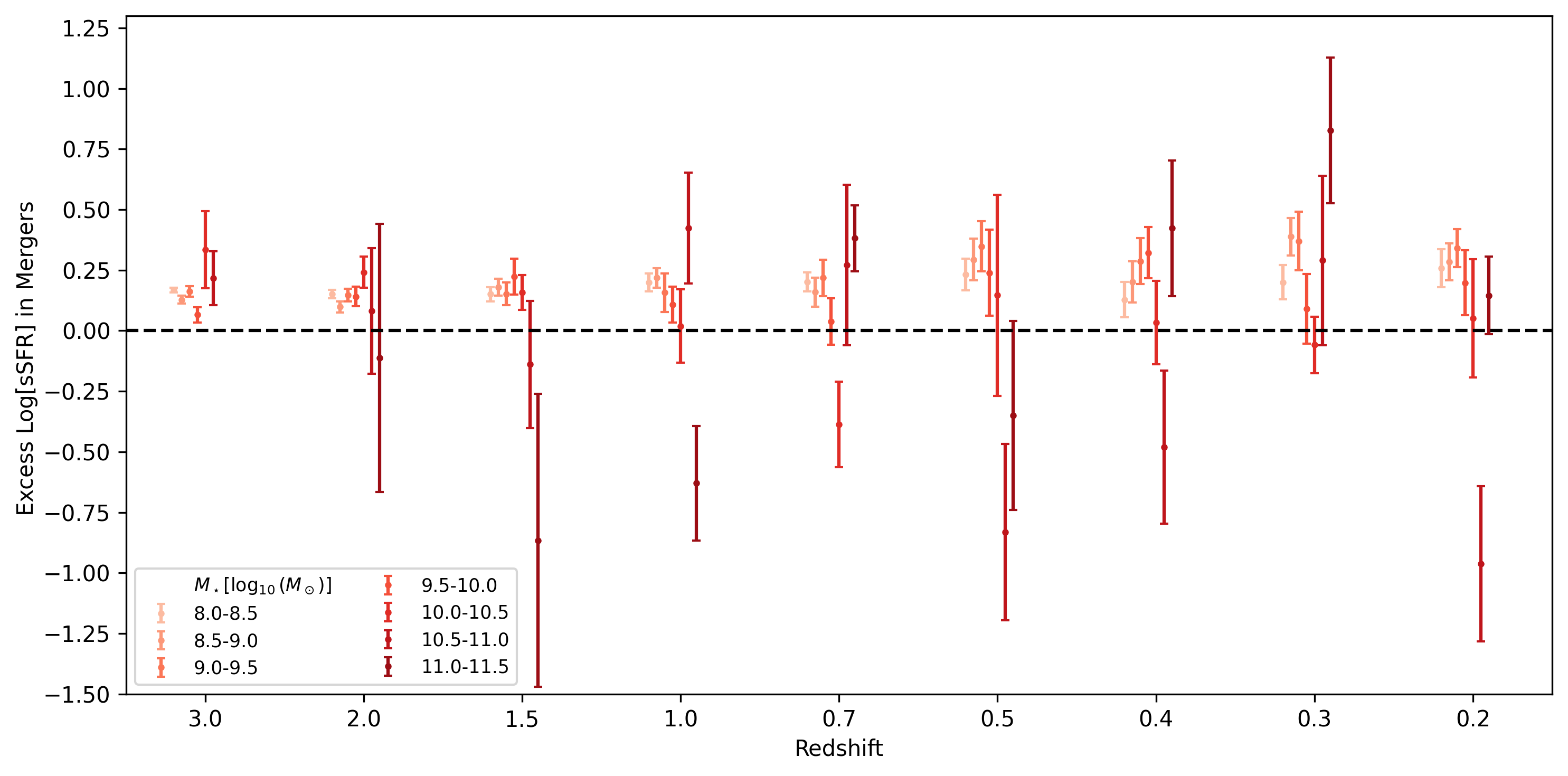}
    \caption{Median excess sSFR ($\log_{10}(\mathrm{sSFR_{mergers}}) - \log_{10}(\mathrm{sSFR_{nonmergers}}))$ and standard error of mergers versus nonmergers at each redshift bin. There is a horizontal offset in each bin added for readability. The color gradient refers to the mass of the galaxies, with lighter orange points for low masses and the darker maroon points for high masses. The black line at zero denotes a level of equal sSFR between mergers and nonmergers. The low mass mergers have higher star formation rates than their nonmerging counterparts, regardless of redshift. As stellar mass increases, the nonmergers sometimes have higher sSFRs, but the difference between median sSFRs of mergers and nonmergers is larger.}
    \label{fig:ssfrexcess}
\end{figure*}

Figure \ref{fig:ssfrexcess} shows the excess sSFRs of mergers versus nonmergers, with all types of mergers included in a single class. 
We define excess sSFR  and excess sBHAR as the ratio of sSFR or sBHAR between mergers and nonmergers ($\log_{10}(\mathrm{sSF(BHA)R_{mergers}}) - \log_{10}(\mathrm{sSF(BHA)R_{nonmergers}}))$.
A positive excess indicates the merging sample has a higher value of sSFR or sBHAR and a negative excess indicates that the nonmerging sample has a higher value.
The mergers at lower stellar masses ($\lesssim 10^{9.5}M_\odot$), at all redshifts in our sample, have higher sSFRs on average than the nonmergers by a factor of $\sim2$.
The mergers at higher masses ($\gtrsim 10^{9.5}M_\odot$) sometimes have excess sSFR, stretching up to $\sim4$ times as high as nonmergers. 
If we only consider galaxies with SFR $> 10^{-11}$ M$_\odot$/yr the negative excess sSFR values come up to hover close to the zero excess line, and if the threshold is SFR $> 10^{-10}$ M$_\odot$/yr the mergers always have excess sSFR.
In TNG50, a stronger trend is seen between merger triggered star formation and stellar mass than between merger triggered star formation and redshift.

\begin{figure}
    \centering
    \includegraphics[width=0.45\textwidth]{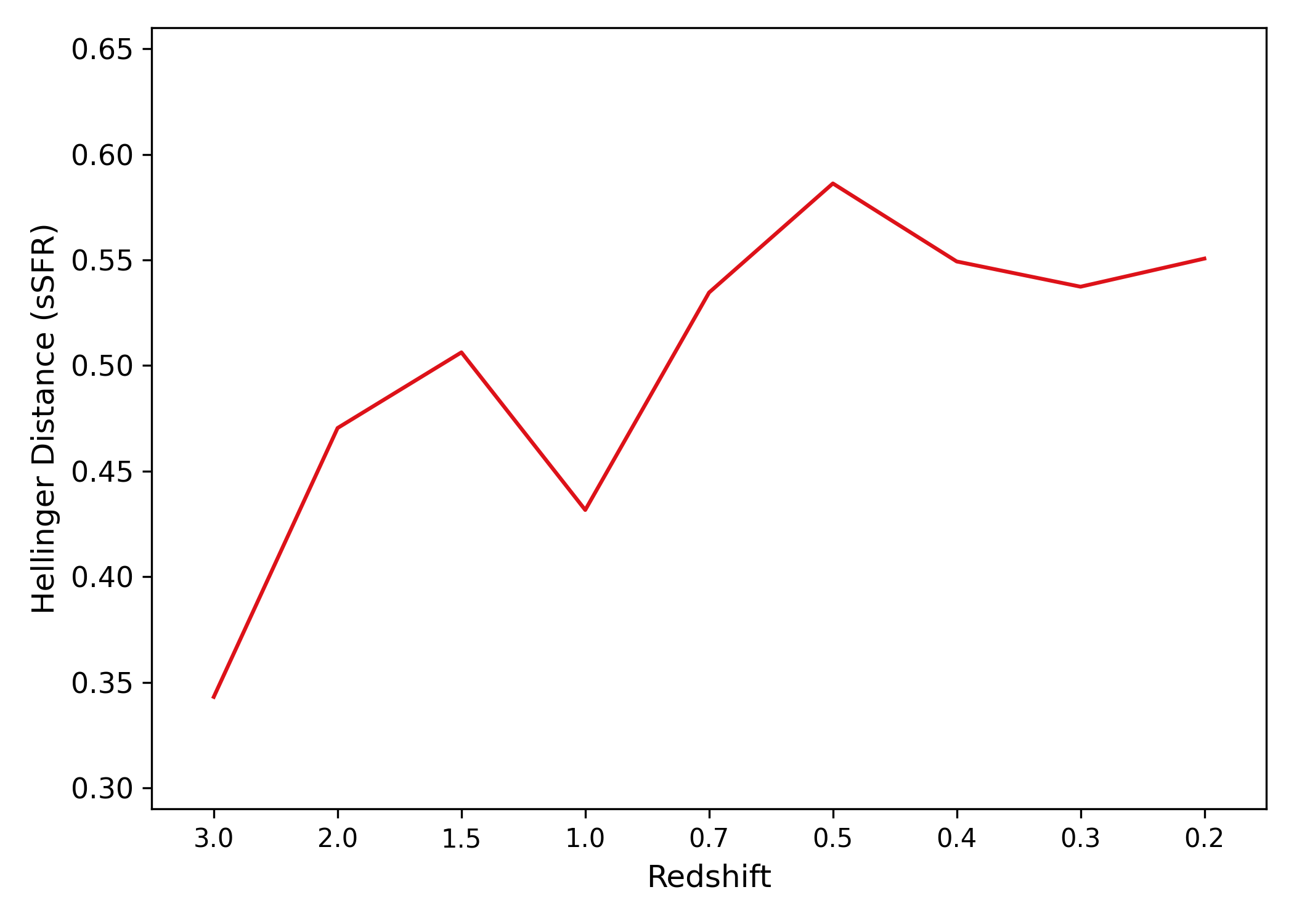}
    \caption{The median Hellinger distance between the mergers' and nonmergers' sSFRs in each redshift bin. A value of 0 indicates statistically similar distributions, and a value of 1 indicates statistically distinct distributions. The mergers and nonmergers become more statistically distinct as redshift decreases at all stellar masses.}
    \label{fig:HellingersSFR}
\end{figure}

To assess the overlap between the distributions of mergers' and nonmergers' sSFRs, we calculate the Hellinger distance in each mass and redshift bin. 
The mass-averaged median Hellinger distance in each redshift bin can be seen in Figure ~\ref{fig:HellingersSFR}. 
The trend of increasing Hellinger distance with decreasing redshift seen in Figure ~\ref{fig:HellingersSFR} is seen at all stellar masses. 
The Hellinger distance is not confirming that the sSFRs are higher in mergers, just that they are a different population than the nonmergers.

\begin{figure*}
    \centering
    \includegraphics[width=\textwidth]{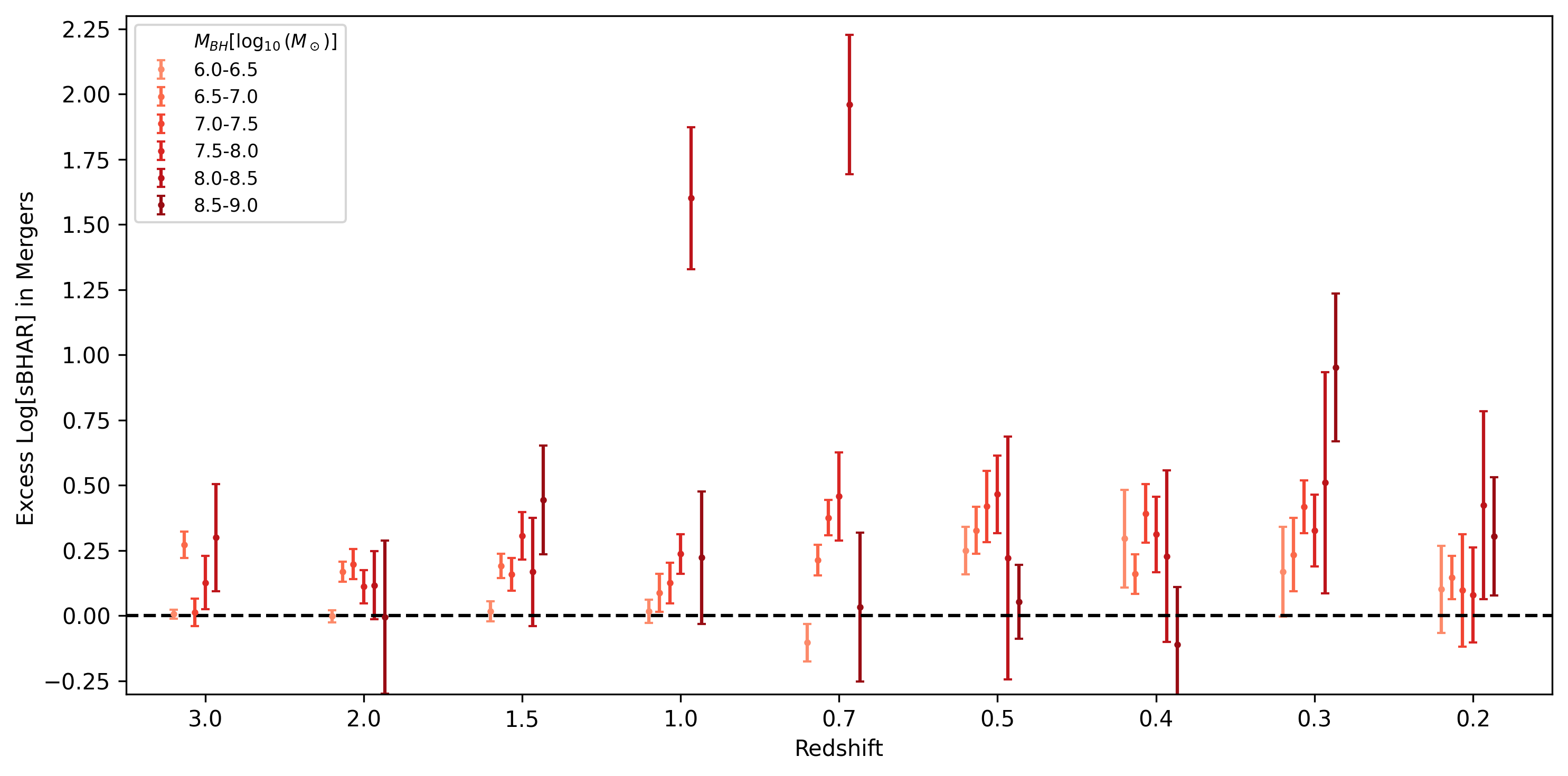}
    \caption{Same as Figure~\ref{fig:ssfrexcess}, except showing black hole mass and sBHAR. The black holes in mergers tend to have higher sBHARs in all mass bins except the smallest, $10^6 < M_\mathrm{BH} < 10^{6.5}$.
    }
    \label{fig:sbharexcess}
\end{figure*}

The black hole accretion activity in merging vs nonmerging host galaxies is shown in Figure ~\ref{fig:sbharexcess}. 
Mergers do generally tend to have excess sBHAR compared to nonmergers, with intermediate redshift bins tending to have the highest excess, with sBHARs anywhere from $\sim 2 - 100$ times as high as nonmergers.
The two bins with extreme excess sBHAR (10 - 100 times), have mergers that maintain their high sBHAR levels before falling to more closely match the nonmergers at lower redshift.

\begin{figure}
    \centering
    \includegraphics[width=0.45\textwidth]{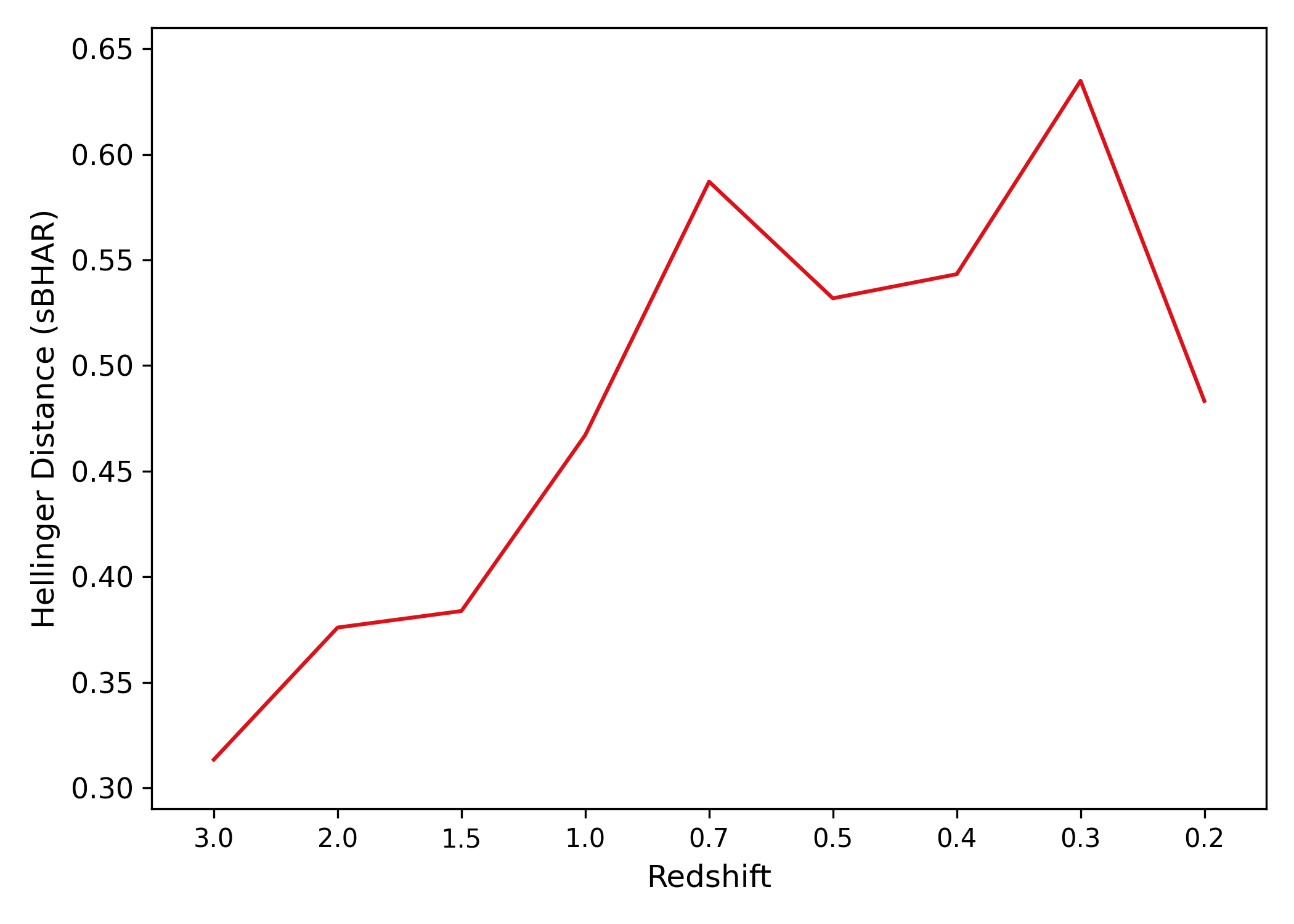}
    \caption{Same as Figure~\ref{fig:HellingersSFR} but for black hole mass and sBHAR. The mergers and nonmergers again become more statistically distinct from each other at any black hole mass as redshift increases. 
    }
    \label{fig:HellingersBHAR}
\end{figure}

To understand how similar the sBHAR distributions are between mergers and nonmergers, we plot the black hole mass-averaged, median Hellinger distance in each redshift bin in Figure ~\ref{fig:HellingersBHAR}. 
We again see that the distance tends to get larger at high masses.
This general positive trend is seen in all black hole mass bins.
The Hellinger distances confirm that the mergers' and nonmergers' sBHARs are statistically different populations. 

\begin{figure*}
    \centering
    \includegraphics[width=\textwidth]{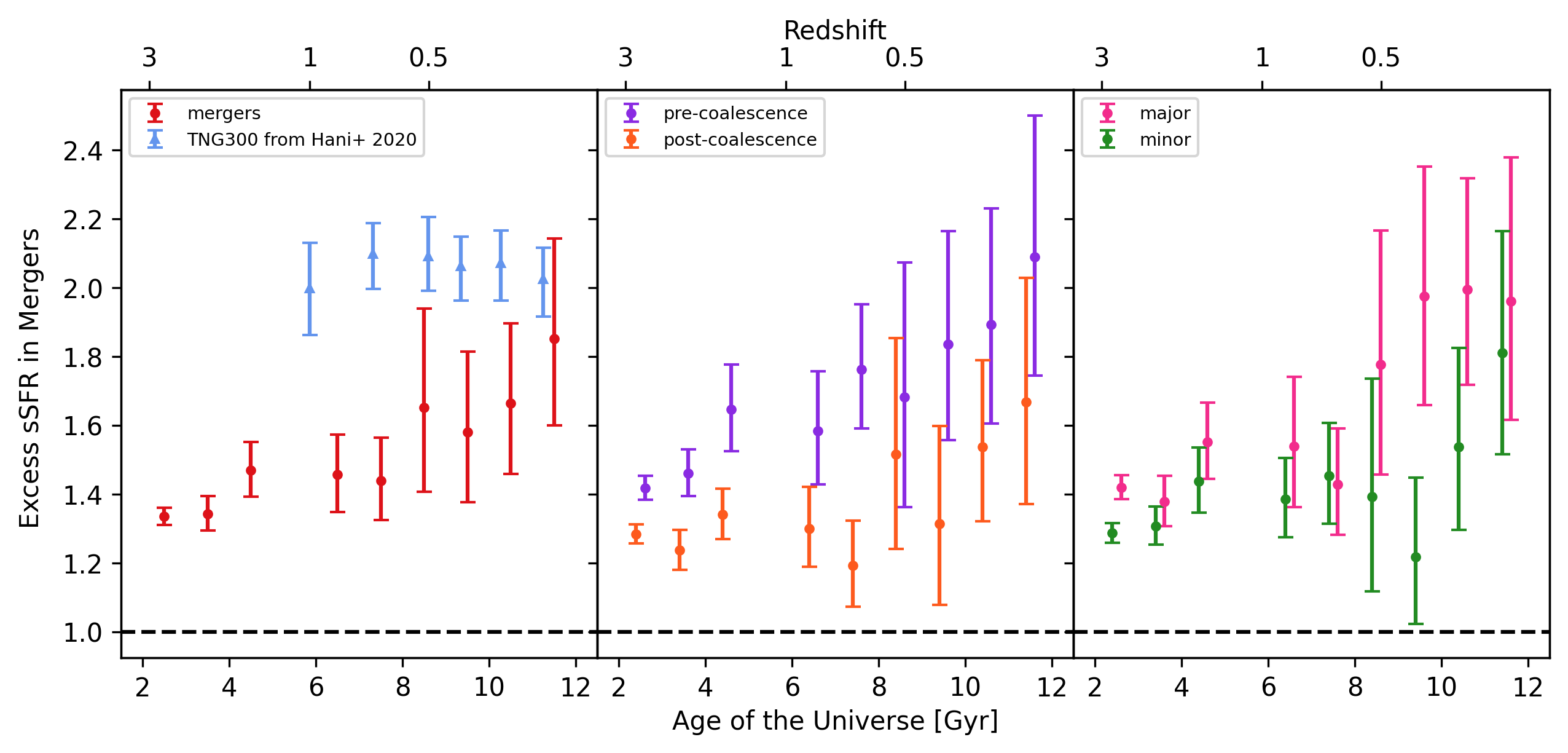}
    \caption{Excess sSFR through cosmic time, defined as the ratio between sSFR for the merging galaxies to the mass-matched nonmerging sample, shown for all mergers and nonmergers (left), mergers split by merger stage (center) and mergers split by mass ratio (right). The error bars are the standard errors in each redshift bin. We compare to excess sSFR reported in \citet{hani_interacting_2020}. Some small horizontal offsets are added for readability. The mergers, in all stages and mass ratios, have excess sSFR compared to the nonmergers at all cosmic times in our study.}
    \label{fig:sSFRvsz}
\end{figure*}

\begin{figure*}
    \centering
    \includegraphics[width=\textwidth]{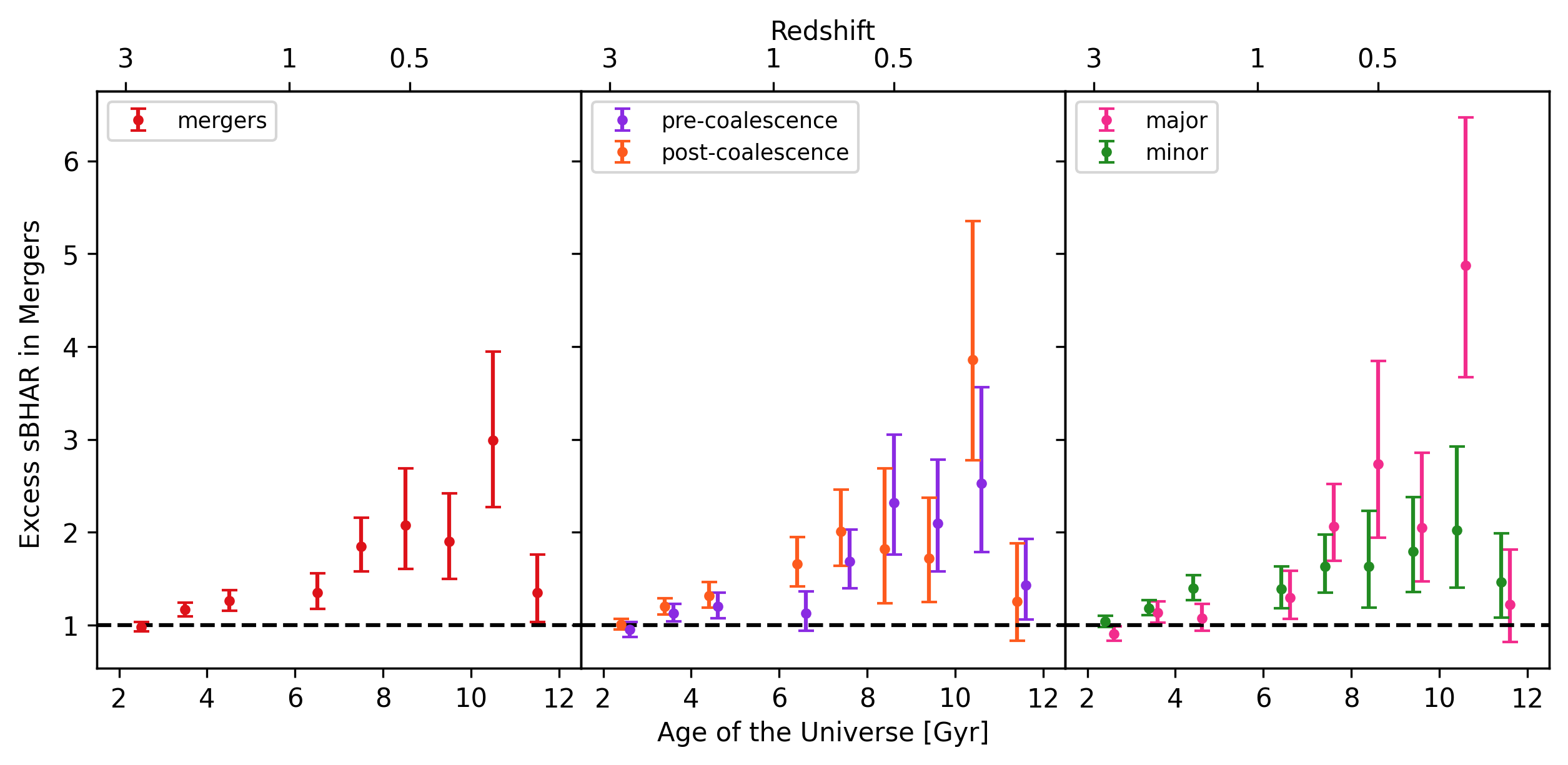}
    \caption{The Excess sBHAR through cosmic time (left), mergers split by merger stage and nonmergers (center) and mergers split by mass ratio and nonmergers (right). The error bars are the standard errors in each redshift bin. Small horizontal offsets are added for readability. Mergers host black holes with excess sBHAR from $0.2 < z < 2$.}
    \label{fig:sBHARvsz}
\end{figure*}

The left hand panels of Figure~\ref{fig:sSFRvsz} and Figure~\ref{fig:sBHARvsz} show the excess sSFRs and sBHARs through cosmic time, with all mergers (of any stellar mass) being compared to their mass-matched nonmerging counterparts at each redshift.
The differences between median sSFR and sBHAR in mergers and nonmergers increase with decreasing redshift.
In TNG50, by $z \sim 3$, mergers already have elevated median sSFRs.
The excess sSFR from \citet{hani_interacting_2020} using TNG300 galaxies is also shown, and while it is also a positive excess, it is larger than the excess we see in TNG50. 
Additionally, \citet{hani_interacting_2020} sees relatively constant excess sSFR from $z = 0.2$ to $z = 1$, while we see an decreasing excess sSFR from $z = 0.2$ to $z = 1$.
The black holes grow similarly in mergers and nonmergers for the first $\sim 4$ Gyr in TNG, with the nonmergers even having higher sBHAR in the highest redshift bins.
The mergers have excess sBHAR and continue to have more and more excess sBHAR as redshift decreases from $0.3 < z < 2$.


\subsection{Merger Stage Has Minimal Impact on sSFR and sBHAR Excess} \label{sec:mergerstage}

We next separate the merging galaxies into pre-coalescence mergers and post-coalescence mergers, and perform the same analysis as Section~\ref{sec:MergersvsNonmergers} in order to determine if merger stage significantly affects the amount of excess star formation or black hole accretion.

Both stages show the same trend in excess sSFR as in Figure~\ref{fig:ssfrexcess}, with low mass galaxies of any stage having excess star formation, and mid to high mass galaxies having more scattered behavior around zero excess sSFR. 
The median Hellinger distances for both merger stages follow the same trend as the overall mergers in Figure~\ref{fig:HellingersSFR}, meaning that both pre-coalescence mergers' and post-coalescence mergers' sSFRs are statistically unique distributions when compared to nonmergers' sSFRs.

When all mass bins are aggregated, mergers of any stage have higher median sSFRs than their mass-matched nonmerging counterparts, as seen in the middle panel of Figure~\ref{fig:sSFRvsz}.
The pre-coalescence mergers often have more excess sSFR than the post-coalescence mergers, but both stages have positive excess values in all redshift bins. 
Separating by merger stage shows that earlier stage mergers may exhibit higher sSFRs, but does not change our overall conclusion that mergers have higher sSFRs than nonmergers at all redshifts in our sample.

Both merger stages also show the same trends in excess sBHAR as in Figure~\ref{fig:sbharexcess}. 
The one exception is the nonmergers have excess sBHAR over the pre-coalescence mergers in a few high redshift and low mass bins. 
Otherwise black holes in both merger stages show more excess sBHAR in mergers, with increases in some mergers of both stages up to factors of $\sim100$ in the high mass bins.
Both stages follow the trend seen in Figure ~\ref{fig:HellingersBHAR}, with the redshift-averaged median distances of both stages increasing as redshift decreases. 

In the middle panel of Figure~\ref{fig:sBHARvsz}, the black holes are split up by merger stage of their host galaxies. 
The behavior in both merger stages is very similar to each other and to that of all mergers on the left panel. 
In the $z = 1$ bin, there is more excess sBHAR in the post-coalescence bin than in the pre-coalescence bin, but as the universe continues to age the black holes in each merger stage behave statistically similarly again.
The merging stage does not appear to be crucial here, causing no unique overall trends in mass or redshift, as the pre-coalescence and post-coalescence mergers behave similarly to each other overall.

\subsection{Merger Mass Ratio Affects sSFR and sBHAR More Than Merger Stage} \label{sec:mergermassratio}
We finally separate the merging galaxies into major and minor mergers to investigate how merger mass ratio may affect sSFRs and sBHARs compared to nonmergers.

The overall trends for both mass ratios are the same as those for all mergers seen in Figure~\ref{fig:ssfrexcess}.
Notably, more extreme positive excess values at high masses appear to come from major mergers.
However, high mass, major mergers also cause some negative excess sSFR \textnormal{in Figure~\ref{fig:ssfrexcess}}, so they are not always producing bursts of star formation, as some galaxies involved in these high mass, major mergers are gas poor.
Minor mergers at low redshift see a decrease in excess sSFR around stellar masses of M$_\star > 10^{9.5}$M$_\odot$.
Major mergers don't see a decrease in excess sSFR until M$_\star > 10^{10}$M$_\odot$.
The Hellinger distances of major mergers are larger than those of minor mergers, but only by $\sim0.1$, and both mass ratios still follow the same trend as Figure~\ref{fig:HellingersSFR}.

Minor mergers do have generally less excess sSFR than major mergers, seen in the right panel of Figure~\ref{fig:sSFRvsz} when all stellar masses at each redshift are in one bin.
While merger mass ratio does impact the behavior, with minor mergers not having the extreme excess sSFR values that major mergers can have, and minor mergers having smaller excess sSFRs on average, mergers of both mass ratios still have higher median sSFRs than nonmergers.
There is a stronger trend with stellar mass than mass ratio of mergers for sSFR enhancement in TNG50, though merger mass ratio is has a larger effect than than merger stage.

The major and minor mergers both follow the same trends in sBHAR as in Figure~\ref{fig:sbharexcess}, with the highest excess sBHAR values coming from the highest black hole mass bins. 
Major mergers are responsible for the largest excess sBHAR values.
Additionally, we note the major mergers at high redshift have some negative sBHAR excess that is washed out by the minor mergers in Figure~\ref{fig:sbharexcess}.
This again is from galaxies in major mergers that are gas poor, and thus the black holes are not accreting material for long periods of time, or have no fuel to accrete.
The Hellinger distances follow the same trends seen in Figure~\ref{fig:HellingersBHAR} with major mergers' distances being larger by $\sim0.1$ than minor mergers, and both mass ratios being distinct populations from the nonmergers.

The right hand panel of Figure~\ref{fig:sBHARvsz} shows the excess sBHAR values for all masses of black holes in one bin at each redshift.
The major mergers have smaller, and even negative, excess sBHARs than minor mergers from $1 \lesssim z \lesssim 3$.
The error bars of these excess values do not overlap at these times, showing that there may be something fundamentally different about the merging process of these mass ratios. 
However, both mass ratios do follow the qualitative trend of the left hand panel of all mergers with excess sBHAR values increasing throughout time. 
Interestingly, the minor mergers never have a clear negative excess sBHAR value, while the major mergers do at early times and again in the lowest redshift bin.
Major and minor mergers behave more differently compared to each other than the two merger stages when examining excess sBHAR, yet the mass of the black hole and whether a host galaxy is merging or not remain stronger influences over merger mass ratio in understanding excess sBHAR.

\subsection{\texorpdfstring{Elevated sSFR and sBHAR in Mergers Lasts for $\sim$1 Gyr}{Elevated sSFR and sBHAR in Mergers Lasts for ~1 Gyr}} \label{sec:timefrommerger}

\begin{figure*}
    \centering
    \includegraphics[width=\textwidth]{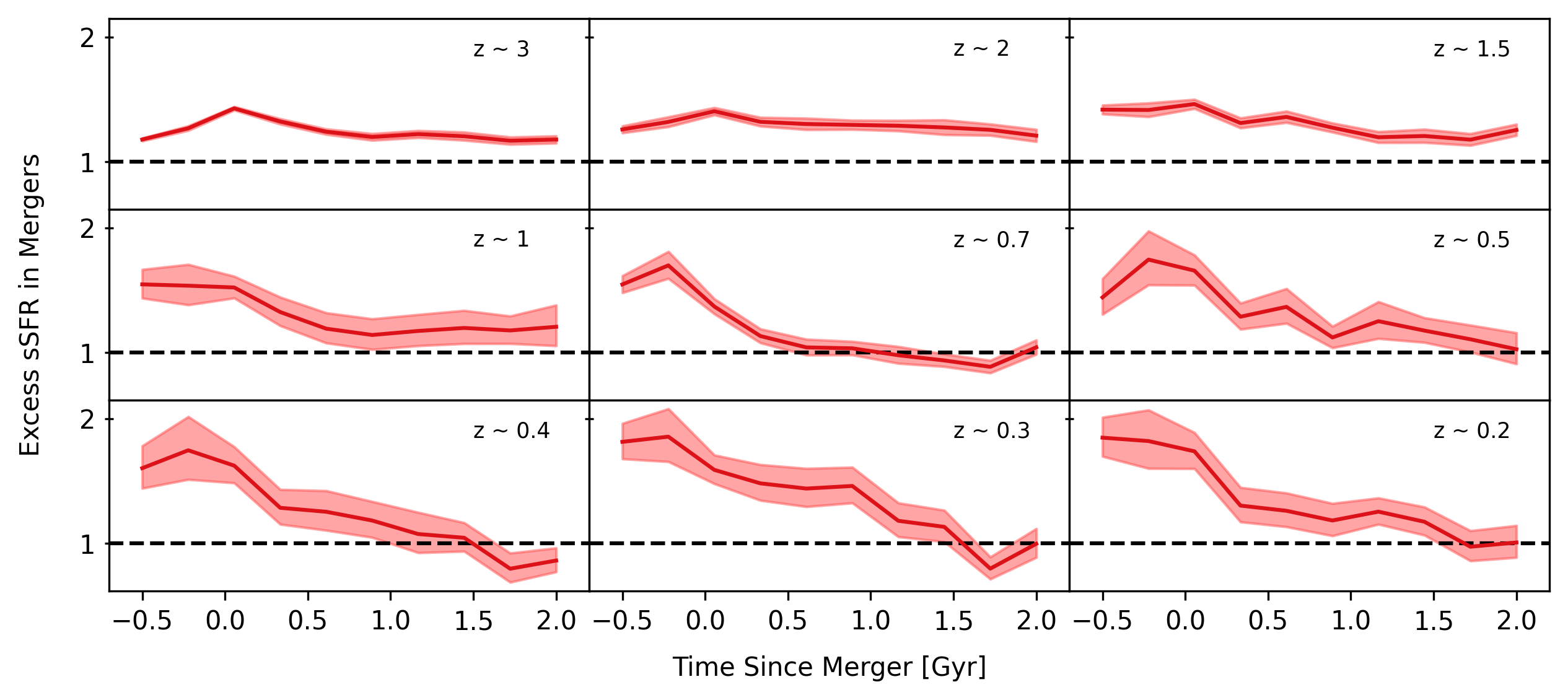}
    \caption{Excess sSFR from 0.5 Gyr before the time of merging to 2 Gyr after in each redshift bin. The shaded regions show the standard error around the median solid line. The dashed line shows an excess of zero: i.e, mergers and nonmergers having the same sSFR. Mergers do tend to have a small spike in sSFR compared to nonmergers at the time of merging, and remain elevated for at least $\sim$1 Gyr after, with high redshift mergers maintaining their excess sSFR for over 2 Gyr.}
    \label{fig:TimeFromMergerSFR}
\end{figure*}

\begin{figure*}
    \centering\includegraphics[width=\textwidth]{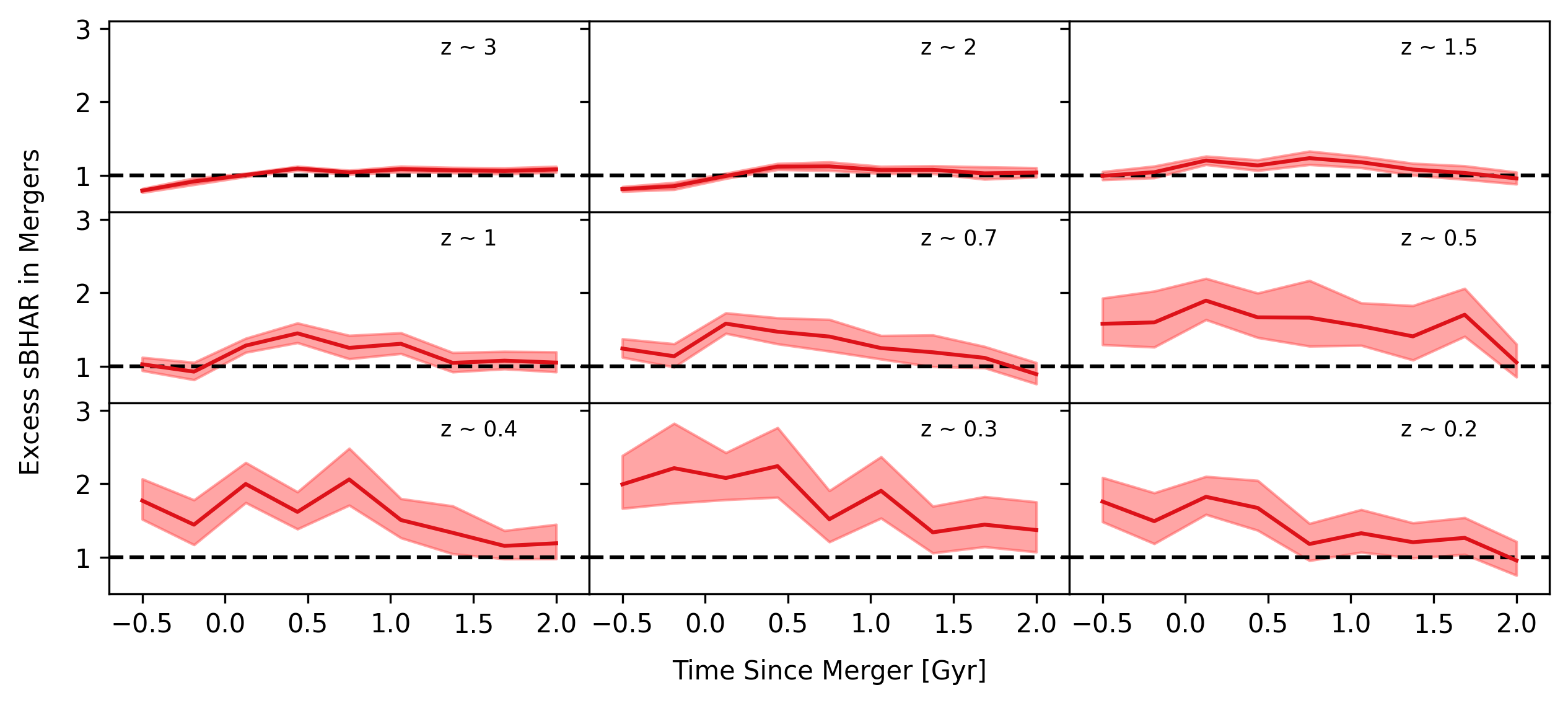}
    \caption{Same as Figure~\ref{fig:TimeFromMergerSFR} but with sBHAR. High redshift black holes in merging host galaxies do not have an excess sBHAR until the time of merging, and though the increase is small, can maintain an excess sBHAR for at least $\sim$1 Gyr. Mid to low redshift galaxies also maintain excess sBHAR for at least $\sim$1 Gyr with a much higher excess value than high redshift mergers.}
    \label{fig:TimeFromMergerBHAR}
\end{figure*}

Lastly, we examine the effect of mergers in TNG50 as a function of time during a merger. 
Figure \ref{fig:TimeFromMergerSFR} shows the median excess sSFR of mergers in each redshift bin over a period of 2.5 Gyr, with $t = 0$ Gyr being the first snapshot where the merger is recognized as a single subhalo. 
Mergers are, on average, more star forming than nonmergers, especially at lower redshifts.
There does tend to be a noticeable peak in sSFR around or a little before $t = 0$, though it is by a factor of a few, not orders of magnitude.
We also note that as redshift decreases, mergers have higher and higher excess values at $t = -0.5$ Gyr.
The mergers continue to have excess star formation for a minimum of $\sim1$ Gyr, and this excess remains for even longer at $1 < z < 3$, around the peak of star formation in the universe.

Figure \ref{fig:TimeFromMergerBHAR} shows the median excess sBHAR and standard error for the same 2.5 Gyr period.
At high redshifts, merging galaxies start with lower sBHAR than nonmerging galaxies, seen by the red medians starting below the dotted black line of equal sBHARs.
The black hole accretion rates do increase around $t = 0$, especially at high redshift. 
The merging galaxies host black holes with higher sBHARs for $\sim1$ Gyr, with a higher excess value (both peak and starting) at low redshfits.
The highest excess sSFR occurs within 0.25 Gyr of $t=0$, while the highest excess sBHAR is often not the first peak in Figure~\ref{fig:TimeFromMergerBHAR}, making it difficult to determine if sSFR peaks first in TNG. 
However, at some redshifts, the sSFR peak is just before $t = 0$, while the sBHAR peaks are always after $t = 0$.

One caveat to note is that by choosing the 50 Mpc box, our galaxies have higher spatial resolution, but we have fewer galaxies overall. 
Because of this, we are unable to impose a limit on how soon the next merger is after the merger at t = 0 in Figures \ref{fig:TimeFromMergerSFR} and \ref{fig:TimeFromMergerBHAR}. 
The nonmerging sample does not merge for 2 Gyr prior to $t = 0$, but may merge later in the 2 Gyr time window shown here. 
Therefore, in some merging systems, some of the elevated star formation or black hole accretion 0.5 Gyr prior to or up to 1 Gyr after could be from another merging event. 
Any elevated sSFRs or sBHARs at $t < 0$ are from a merger even if it is not from the merger we focus on.
Additional mergers occur later in the time window at similar rates between the selected mergers and nonmergers (based on classifications at $t=0$), so this does not affect our conclusions.

\subsection{Mergers Have Higher Gas Fractions Than Nonmergers}\label{sec:Gas}

\begin{figure*}[htbp]
    \centering
    \includegraphics[width=\textwidth]{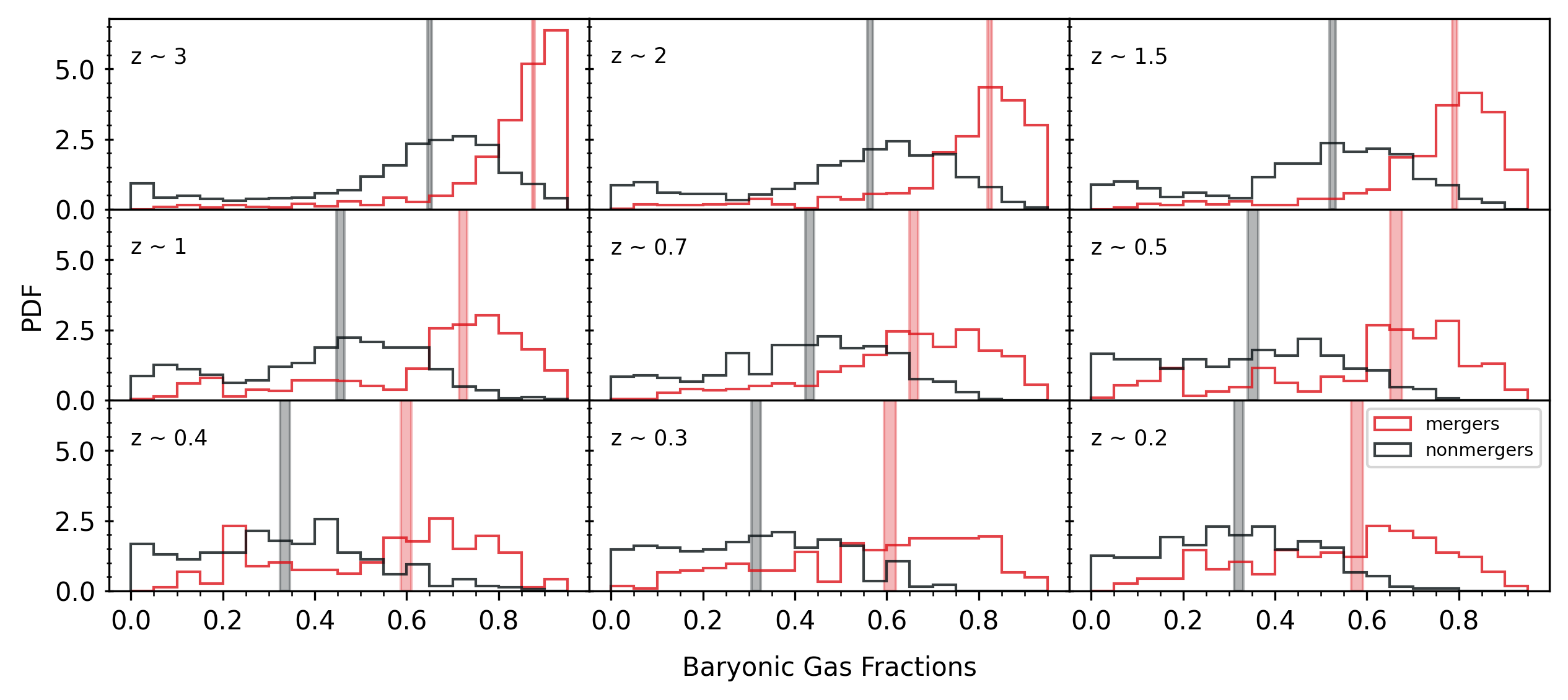}
    \caption{The baryonic gas fractions of mergers (red) and nonmergers (black) at each redshift bin. The shaded regions encompass the median of each distribution and one standard error. This higher gas fraction among mergers can explain why the mergers often have more star formation than the nonmergers.}
    \label{fig:gasfrac}
\end{figure*}

\begin{figure*}[htbp]
    \includegraphics[width=\textwidth]{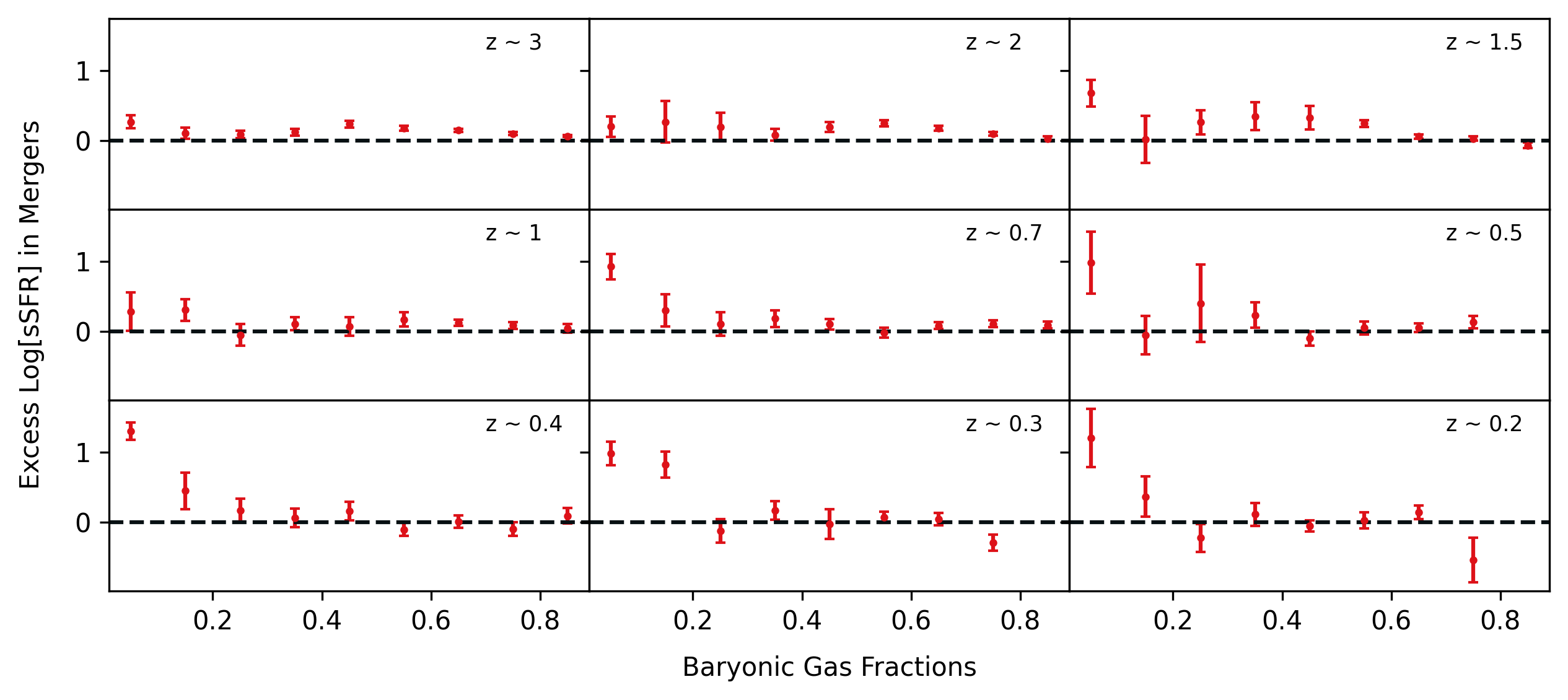}
    \caption{The excess sSFR of mergers to nonmergers at each redshift bin as a function of gas fraction. \textnormal{At the highest redshifts, we see little significant elevation in sSFR at any gas fraction. At $z \lesssim 1$, we start to see higher sSFR enhancements at gas fractions of less than about 20\%, in agreement with the predictions from idealized binary merger simulations.} 
    }
    \label{fig:sfe}
\end{figure*}

In order to explain why the merging galaxies often have more star formation than the nonmerging galaxies we look into their gas content. 
Following \citet{hani_interacting_2020}, we define the gas fraction as
\begin{equation}
    f_{\mathrm{gas}} \equiv \frac{\sum\limits_{\mathrm{prog}}M_{\mathrm{gas}}}{\sum\limits_{\mathrm{prog}}M_{\mathrm{gas}} + \sum\limits_{\mathrm{prog}}M_\star}
\end{equation}
where $\sum\limits_{\mathrm{prog}}M_{\mathrm{gas}}$ is the sum of the progenitors' gas mass and $\sum\limits_{\mathrm{prog}}M_\star$ is the sum of the progenitors' stellar mass. 
As with our previous calculations, the quantities are taken within twice the stellar half mass radius. 
This does not take into account the temperature of the gas, so the gas fractions shown here are not only cold gas that would form stars, but all gas in the system.
Therefore, this is not directly comparable with cold gas fractions in observations.

Figure \ref{fig:gasfrac} shows the baryonic gas fractions in both mergers and nonmergers. 
The two populations follow the same general trends through time with the galaxies of both classes tending to have more gas than stars at high-$z$, and gas fractions becoming lower as the galaxies form stars and become more gas poor over cosmic time towards low $z$.
The merging population on average has a higher gas fraction than the nonmergers throughout cosmic time, shown by the medians in Figure \ref{fig:gasfrac}.
As galaxies merge, material is driven towards the center.
Therefore, it is not surprising that galaxies undergoing a merger have more gas relative to stars in the nuclear region as gas from their surrounding environments is forced inwards.

\textnormal{We finally examine sSFR as a function of gas fraction to determine the cause of the mergers' enhanced sSFR. 
This relationship is shown in each redshift bin in Figure ~\ref{fig:sfe}.
The mergers tend to have more excess sSFR at gas fractions below 20 percent. 
This trend is less pronounced in the higher redshift bins, 
but becomes quite prominent from $0.2 < z < 1.5$, when disks in TNG50 become kinematically cooler \citep{pillepich_first_2019}.
In agreement with predictions from idealized binary merger simulations, this shows that mergers can drive strong gravitational torques and hence excess star formation more effectively when disks are thin and gas fractions are low \citep{robertson_molecular_2008}.}

\section{Discussion} \label{sec:discussion}
\subsection{Comparison to Other Works -  Simulations}

\textnormal{Our current theoretical understanding of the impact of mergers on SFR and BHAR enhancements was developed through a generation of simulations of binary mergers of idealized, isolated galaxies. \citet{mihos_triggering_1994} and \citet{mihos_gasdynamics_1996} showed that mergers produce strong torques that drive gas into galactic nuclei, where the increased gas density leads to more efficient star formation (due to the super-linear slope of the comonly adopted Kennicutt-Schmidt type star formation recipe). These early works also showed that major mergers drive much stronger inflows and hence more prominent star formation enhancements than minor mergers \citep{cox_effect_2008}. The SFR enhancement also depends on the orbit of the merging galaxy, and the presence of a bulge can stabilize the disk against the merger-driven instabilities, leading to smaller SFR enhancements \citep{cox_effect_2008}. The merger-induced inflows also lead to enhanced accretion onto nuclear supermassive black holes \citep{springel_black_2005}, although the peak in BHAR typically occurs at a later stage in the merger than the peak in SFR \citep{hopkins_dynamical_2012}. The loss of angular momentum that leads to both enhanced SF and BH accretion is largely driven by internal gravitational torques of the stellar component on the gas, resulting from asymmetries induced by the merger. Thus, we expect a dependence of SFR/BHAR enhancement on the initial gas fraction, and in particular expect galaxies with very high gas fractions to (somewhat counterintuitively) have lower SFR/BHAR enhancements due to forming more stars in their isolated phase before the merging event \citep{springel_formation_2005,hopkins_effects_2009,robertson_molecular_2008}. Conversely, if the gas fraction in the progenitors is extremely low (``dry'' mergers), we also expect little impact of the merger on the SFR/BHAR. 
}

Previous work with the larger TNG volumes has shown that merging galaxies have both enhanced sSFRs at small galaxy separations \citep{patton_interacting_2020}, and in the post-coalescence phase \citep{hani_interacting_2020}. 
These sSFRs are enhanced by a factor of $\sim 1.5 - 2$, which while larger than the enhancements shown in this work, do match our qualitative conclusions shown in Figures \ref{fig:ssfrexcess}, \ref{fig:sSFRvsz} and \ref{fig:TimeFromMergerSFR} that mergers have excess sSFR. 
Importantly, these works investigated mergers out to $z = 1$, while our work extends out to $z=3$. 
Comparing only $z < 1$, where our samples overlap, and using the peak of excess sSFR throughout the merger process from Figure \ref{fig:TimeFromMergerSFR}, we also find that the peak in excess sSFR is around a factor of 2.

Our results additionally agree with \cite{hani_interacting_2020} that mergers of all mass ratios experience higher sSFRs with major mergers experiencing the largest enhancement at $z < 1$. 
However, \cite{hani_interacting_2020} also find that their sSFR enhancement fades after $\sim 0.5$ Gyr, which is in contrast to our results shown in Figure \ref{fig:TimeFromMergerSFR} that mergers have excess star formation for $\geq 1$ Gyr. 
Important to note is that due to the larger box size of TNG300, and grouping all galaxies from $0 < z < 1$ into one bin, \cite{hani_interacting_2020} is able to chose a merging sample that does not merge again during the examined time frame. 
With the smaller box size and narrower bin size, we are unable to control for that aspect without encountering limitations from small number statistics.
Multiple mergers in the 2.5 Gyr timeframe are likely inflating how long our merger-driven sSFR spike lasts.

At a similar redshift range in the cosmological simulation Horizon AGN, \cite{kaviraj_galaxy_2015} finds that major mergers only increase star formation by $\sim 20-40\%$, and therefore are not driving the stellar mass growth of the universe. 
SIMBA shows SFR enhancements of $\sim2-3$ in mergers from $0 < z < 2.5$ \citep{rodriguez_montero_mergers_2019}, and sees the same trend we do of high mass, low redshift mergers having lower sSFRs.
We find in Sections \ref{sec:MergersvsNonmergers} and \ref{sec:timefrommerger} that mergers may have a 40 - 50\%, or factor of 1.5, increase in star formation.
We agree with \citet{rodriguez_montero_mergers_2019} that at early times, mergers are driving less excess sSFR, and that as the universe reaches cosmic noon and beyond, mergers drive more excess sSFR.
In EAGLE from $0 < z < 1$, the excess sSFR is a factor of $\sim1.5$, similar to what we find but half of that found in TNG100 \citep{patton_interacting_2020}.

\textnormal{The high resolution TNG50 box includes a large population of disk dominated galaxies with stellar masses of $10^{9-10.5}M_\odot$.} 
Thick disks are more stable to merger-driven torques than thin disks due to being kinematically hotter \citep{quinn_heating_1993}.
Gaseous disks in TNG50 are kinematically hotter at $z \gtrsim 1.5$ \citep{pillepich_first_2019}, meaning mergers of disk galaxies at $z \gtrsim 1.5$ are less likely to see large sSFR enhancements, as their thick disks stabilize them. 
This may explain why we see a larger sSFR enhancement for mergers at low-\z{} in our sample than at high-\z{}.

The gas fractions in the TNG universe are presented in \citet{pillepich_simulating_2018, pillepich_first_2019}, and are much lower than the gas fractions presented in this work. 
However, those are gas fractions based on the total dynamical mass of the galaxy, and thus are expected to be smaller. 
The same trend of significantly higher gas fractions at high-$z$ than at low-$z$ appears in the dynamical gas fractions as in our gas fractions.
\textnormal{The result that nonmerging galaxies in TNG300 have enhanced star formation compared to post mergers at $f_{gas} < 0.1$ from \citet{hani_interacting_2020} contradicts our result in Section~\ref{sec:Gas}.
However, their sample only includes galaxies with $10^{10} < M_\star/M_\odot < 10^{12}$, and thus is more likely to include bulge dominated galaxies than our sample, which stretches to lower masses.
}

\textnormal{Bulges can stabilize disks against radial gas inflows and starbursts \citep{mihos_triggering_1994, mihos_gasdynamics_1996}.
Therefore, the disagreement in low gas fraction sSFR enhancement could be due to bulge-dominated galaxies making up a smaller fraction of our overall sample than in \citet{hani_interacting_2020}.
With our sample stretching to lower masses, we have a larger fraction of disk galaxies.
Alternatively, this discrepancy could be due to the resolution, discussed in depth in Section \ref{sec:resolution}. 
The gas in TNG50 reaches higher densities than the gas in TNG300.
As mergers funnel gas towards the centers of galaxies, the gas densities rise, forming new stars.
If TNG300 does not see the full effect of high gas densities through the merging process, the merger-driven star formation may be underestimated compared to that in TNG50.}

\citet{byrne-mamahit_interacting_2022} find that in TNG100, mergers are more likely than nonmergers to host high luminosity AGN, but AGN are still rare in the merging sample. 
Our results on TNG50 agree, with the range of sBHARs in mergers being higher than that of nonmergers. 
The TNG100 black holes have sustained sBHAR enhancements after a merger \citep{byrne-mamahit_interacting_2022, byrne-mamahit_interacting_2024}, which we see in Figure \ref{fig:TimeFromMergerBHAR}. 
This paper has a wider redshift range and lower stellar mass cutoff for the subhalo than \citet{byrne-mamahit_interacting_2022, byrne-mamahit_interacting_2024}, which studied the range $0 < z < 1$.
TNG100 AGN can be triggered up to 1.5 Gyr before the ``merging snapshot'' \citep{byrne-mamahit_interacting_2024}, which we see evidence for at $z \lesssim 0.7$ in TNG50.

Black hole activity likely depends both on merger stage, but also mass ratio, because of the shorter dynamical times in more massive systems. 
\citet{capelo_growth_2015} finds black hole activity peaks at smaller separations, which would be our post-coalescence mergers.
Our data do not show a definitive overall trend with merger stage for black hole activity, but instead it generally seems to be black hole mass and merger stage dependent.

\subsection{Impacts of the TNG Model: Black Holes and Resolution} \label{sec:resolution}
Though our results and those of TNG as a whole match qualitatively between different resolutions, we note here what effects increasing the spatial resolution and the time between snapshots have on the quantities we calculate in this work compared to zoom in simulations. 
We additionally discuss the black hole model.

Increasing the resolution increases many values of galaxy population statistics, stemming from the increase in stellar mass at fixed halo mass \citep{pillepich_simulating_2018}.
Specifically relevant is that the black hole masses tend to be larger \textnormal{in TNG50 than in TNG100} in for a given halo mass for galaxies of stellar mass $\lesssim 10^{10}$M$_\odot$, which is the majority of our galaxies in this paper \citep{pillepich_simulating_2018}.
Additionally noted in \citet{pillepich_simulating_2018}, a direct consequence of increasing spatial resolution in all simulations (not just TNG) is more gas being available to form stars as the simulation can probe higher density regions.
This leads to elevated star formation rates.
However, feedback from galactic winds can stabilize some of this star formation resolution effect, since high star formation rates will also lead to more galactic wind feedback.

Though we did see a merger-driven rise in star formation and black hole accretion in Section~\ref{sec:timefrommerger}, TNG50 has more time in between snapshots than some isolated galaxy merger simulations, and thus even stronger peaks might be washed out by the coarse snapshot spacing.
Isolated merger simulations show that these SFR peaks may only last for $\sim 500 $  Myr \citep{mihos_triggering_1994, di_matteo_energy_2005, moreno_interacting_2019}.
The difference in time between when SFR and BHAR activity peak can be as small as 0.01 - 0.1 Gyr \citep{hopkins_dynamical_2012}.
Work has also examined how small separations must be in order to see dual AGN activity, but due to TNG's black hole placement and efficient quenching, combined with the snapshot separation discussed above, we may not see activity like dual AGN activating at different times in a merger such as in \citet{petersson_starburst_2023}.

Though the black hole feedback models in TNG are tuned so that the galaxy population statistically matches observed populations, there are of course some oversimplifications.
The black holes in TNG move to the center of the potential well immediately, when in reality black holes would take time to migrate to the center of the potential during a merger.
With FIRE physics, \citet{catmabacak_black_2022} showed that black hole dynamics does not affect black hole masses, but does affect the $M_{BH} - M_\star$ scaling relation. 

TNG's black hole kinetic feedback mode is incredibly efficient at quenching galaxies. 
At $M_{BH} \simeq 10^8$M$_\odot$, most galaxies quench rapidly, \textnormal{as the mass accretion rates and black hole masses cause them to} enter the kinetic feedback mode.
\textnormal{The kinetic mode removes fuel from the center of these galaxies.}
This effect may be influencing our results when we see dips in sBHARs at high masses. 
This effect of course also impacts sSFRs, but with our mass binning scheme it should be affecting the mergers and nonmergers equally, assuming galaxies with similar stellar masses have similar black hole masses, and thus is minimal when measuring excess sSFR.
Due to the small box size of TNG50, there are not enough merging galaxies with high black hole masses to draw statistically significant conclusions on the effects of black holes that cross this effective quenching mass threshold due to a merger.

\subsection{Comparison to Other Works - Observations}
\textnormal{Comparing results from simulations to observationally selected samples of merger candidates is extremely challenging, due to differences in merger selection methods as well as uncertainties in observational tracers of SFR and BHAR. 
In this section, we discuss what previous observational studies have found regarding merger driven SFR and BHAR/AGN activity.}

\subsubsection{Observations of Merger Driven Star Formation}

At low-$z$, with SDSS, observational studies find enhanced SFRs in mergers, especially in major mergers (e.g., \citealt{ellison_galaxy_2008, patton_galaxy_2011,  scudder_galaxy_2012, patton_galaxy_2013, luo_connections_2014, kaviraj_importance_2014, willett_galaxy_2015}).
\citet{scudder_galaxy_2012} notes that order-of-magnitude enhancements are rare, which agrees with our findings in Section~\ref{sec:mergermassratio}.
Our work also agrees with \citet{willett_galaxy_2015}, who find that major and minor mergers boost star formation at a similar level (Figure~\ref{fig:sSFRvsz}). 
Stretching to higher redshifts, with \textit{HST} data, \citet{barrows_observational_2017} showed that higher sSFRs are more likely to occur in mergers, and that sSFRs fall as separation decreases, meaning star formation peaks earlier in the merging process. 
Most work studying the impact of mergers on SFR is at low-$z$ due to the difficulty of identifying more distant mergers.

\textnormal{We now compare our sSFR excess and gas fractions to those measured observationally.
Among local galaxies where it is possible to take deep HI observations, gas fractions tend to decrease as stellar mass increases, and bluer (in \textit{NUV-r} color; a proxy for sSFR) galaxies tend to have higher gas fractions \citep{brown_effect_2015}.
The positive sSFR - gas fraction relationship holds out to higher redshifts, with \citet{tacconi_phibss_2018} showing at fixed stellar mass and redshift, out to $z = 4$, the gas fraction increases as sSFR increases.
\citet{combes_gas_2013} used CO line detections to measure gas fractions out to $z  = 1$, finding that gas fractions are a factor of $3\pm{1}$ higher at $z = 1$ than at $z = 0$.
These trends match what we find in Figure~\ref{fig:gasfrac}, where the higher redshift bins (which contain more low mass galaxies) have higher median gas fractions.}

\textnormal{To examine the gas content of galaxy mergers, \citet{violino_galaxy_2018} analyzed early stage merger and \citet{ellison_enhanced_2018} analyzed late stage mergers of nearby galaxies.}
\citet{violino_galaxy_2018} calculates gas fractions ($f_{gas} = M_{H_2}/M_\star$) in SDSS close pairs. 
They show that at a fixed stellar mass, close pairs have higher gas fractions than isolated counterparts. 
The close pairs also have enhanced sSFR compared to the nonmergers.
We see both of these effects in our results.
Additionally, they study the gas depletion time, and find the close pairs have a shorter depletion time, and are thus more efficient at turning gas into stars. 
They attribute this to the merging process, as the interaction causes some of the gas to shift into a denser phase.

\textnormal{At fixed stellar mass, postmergers in SDSS (also observed with Arecibo) have elevated global atomic hydrogen fractions ($f_{gas} = \textrm{log}[M_{HI}/M_\star]$) compared to a control sample of galaxies from xGASS, a survey of $\sim 1200$ galaxies ($ 10^9M_\odot < M_\star < 10^{11.5}M_\odot$) with 
a $f_{gas}$ lower limit of $\sim 2$ percent \citep{catinella_xgass_2018, ellison_enhanced_2018}. 
The median HI gas fraction is 0.3-0.6 dex larger for the postmerger sample than the control xGASS sample.
They attribute this primarily to combining galaxies' gas reservoirs.
They simulate mergers' HI gas fractions at a given stellar mass by combining two control galaxies from the xGASS sample and calculating what the gas fraction would be of this ``merger".
Since lower mass galaxies have higher gas fractions, and vice versa, combining two lower mass galaxies to form a galaxy of stellar mass $M_\star$ results in a gas fraction that is higher than that of an isolated galaxy of stellar mass $M\star$.}

\textnormal{Comparing galaxies with similar initial SFRs can help determine whether a merger or instability is responsible for SFR enhancement.}
Interestingly, when \citet{violino_galaxy_2018} match the close pairs in both stellar mass and SFR, \textnormal{so that quenched galaxies are only compared to other quenched galaxies, and star forming galaxies only compared with star forming galaxies,} these merger enhancements disappear. 
Therefore, after matching in both stellar mass and SFR, \citet{violino_galaxy_2018} finds instabilities may cause the same effects on the gas that mergers do. 
\citet{ellison_atomic_2019} finds the same when comparing AGN host galaxies, though only with nonmerging AGN host galaxies. 
The AGN hosts have higher HI gas fractions when matched only with stellar mass, but if the control sample additionally includes SFR, the gas fractions are consistent with the controls.
Both of these studies are with low-$z$ galaxies, which enables higher quality spectra with which to make these measurements. 
We are limited by the box size in TNG50 to match our merging and nonmerging samples in SFR in addition to stellar mass.
\textnormal{The TNG50 box contains so few galaxies that matching with SFR causes our mass match thresholds to widen considerably, ultimately giving a poorer overall matched nonmerging galaxy to each merging galaxy.
We include all gas particles in TNG, so the fraction itself does not directly correspond to the observed HI or H$_2$ fractions in \citet{ellison_enhanced_2018} and \citet{violino_galaxy_2018}, thus a true comparison is not possible.}

\subsubsection{Observations of Merger Driven Black Hole Activity}

Observations disagree on the extent to which mergers trigger AGN, with general agreement that mergers can, but do not always, trigger AGN (e.g., \citealt{georgakakis_host_2009, cisternas_bulk_2011, kocevski_candels_2012, rosario_mean_2012, villforth_morphologies_2014, mechtley_most_2016, villforth_host_2017, barrows_census_2023}).
The mass ratio of the merger, redshift of the system, AGN luminosity, and observed wavelength are all factors which affect the strength of the merger-AGN connection. 
This paper finds that in TNG, the mass of the galaxy and black hole are more closely related to black hole activity than merger stage or mass ratio.
There is clearly a redshift evolution, especially for high mass galaxies, in how elevated the sBHARs of mergers are.
Observations also find that mergers may be more important for triggering AGN at $z \lesssim$ 1 \citep{kocevski_candels_2012}.

Many works find that major mergers dominate the host galaxy population of high luminosity AGN (e.g., \citealt{treister_major_2012}, \citealt{comerford_offset_2014}, \citealt{comerford_merger-driven_2015}, \citealt{glikman_major_2015}, \citealt{ellison_definitive_2019}).
On the other hand, \citet{simmons_moderate-luminosity_2012} shows that minor mergers or secular processes could be driving black hole growth at $z \sim 2$, which is a higher redshift than many of the papers showing major mergers dominating luminous AGN. 
Our results show that major mergers are important for elevating sBHARs, but the trend with galaxy mass is stronger. 
The contribution of minor mergers is non-negligible, and in fact very significant at high masses, and at early times. 
At $z \sim 3$, nonmerging galaxies have equal or more sBHAR than merging galaxies, implying that mergers are not needed for high sBHARs at early times (Figure~\ref{fig:sBHARvsz}).

Our results show that mergers in TNG50 do have higher sBHARs than nonmergers at $z \lesssim 3$, and that major mergers do tend to have higher sBHARs than nonmergers and minor mergers in many cases. 
However, we also see that minor mergers are non-negligible, and also tend to have higher sBHARs than nonmergers.

\textnormal{Depending on the wavelength range observed, studies target different types of AGN.
Radio-loud AGN are detected from their emission from relativistic jets, producing large lobes detected at radio wavelengths (e.g., \citealt{dunlop_quasars_2003, ramos_almeida_optical_2011,chiaberge_radio_2015}).
AGN that are detected only in the IR tend to be dust-obscured, luminous AGN, and are found more often in mergers than nonmergers \citep{satyapal_galaxy_2014, donley_evidence_2018, ellison_definitive_2019, villforth_complete_2023}.
Optically detected AGN are found through broad emission lines in spectra (e.g., \citealt{ hong_correlation_2015,villforth_host_2017, marian_major_2019, marian_significant_2020}). 
However, dust can often obscure these broad emission lines, so these detections are often combined with IR detections.
AGN detected in the X-ray only are likely lower luminosity AGN with less obscuration \citep{eckart_comparison_2010, donley_identifying_2012}.}

\textnormal{Most observational AGN studies only consider major mergers, and are therefore also not directly comparable to our sample that includes mergers with stellar mass ratios down to 1:10.
Many studies that show no preference for AGN to reside in major mergers do note that minor mergers may play a role in triggering AGN activity.
Merger identification methods can play a role in determining the merger - AGN connection as well, as close pairs may be missing AGN triggered at a later merger stage, and postmergers may be overestimating the merger incidence, due to not counting early stage mergers that have not yet triggered AGN activity.}

\textnormal{Studies with X-ray-selected AGN often find no preference for merging galaxies to host AGN \citep{georgakakis_host_2009, cisternas_bulk_2011, bohm_agn_2013, kocevski_candels_2012, villforth_morphologies_2014, hewlett_redshift_2017}. 
The exception is \citet{koss_merging_2010}, who do find a higher incidence of AGN among galaxies with visual signs of a merger.
\citet{koss_merging_2010} is the lowest redshift study, only looking at $z < 0.05$, while the other works stretch to a maximum redshift of $z = 2.5$ in \citet{kocevski_candels_2012}.
Mergers may be more crucial to triggering AGN activity in the local universe as galaxies' cold gas fractions decrease through cosmic time \citep{santini_evolution_2014, popping_inferred_2015}. }

\textnormal{
AGN detected in the optical and infrared follow the same trend as X-ray detected AGN with no AGN - merger connection at cosmic noon, but with mergers becoming significant drivers of AGN activity at low-\z{}. 
Mergers are no more likely than nonmergers to host AGN in \HST{} observations of galaxies at $z \sim 0.6$ \citep{villforth_host_2017}, or at $z \sim 2$ \citep{mechtley_most_2016, marian_major_2019}.
However, similarly to the X-ray AGN discussed above, in the local universe there is evidence that major mergers do trigger AGN.
\citet{ellison_definitive_2019} combines optical and IR AGN detections at $z \sim 0 $ and finds that while optically selected AGN are just as likely to be found in mergers and nonmergers, the IR selected AGN are at least twice as likely to be hosted by a galaxy showing signs of a recent interaction. 
\citet{hong_correlation_2015} and \citet{marian_significant_2020} support the idea that mergers trigger AGN locally and find a clear preference for AGN to reside in merging hosts at $z < 0.2$. 
}

\textnormal{Meanwhile, studies do find evidence of radio loud AGN existing preferentially in mergers (e.g., \citealt{mclure_sample_2004, chiaberge_radio_2015}), though the sample sizes are small.
\citet{chiaberge_radio_2015} cites black hole - black hole mergers that occur during a galaxy merger as a likely cause of this radio loud AGN and merger connection. This is because a black hole merger may spin up the resulting black hole, and radio loud AGN are associated with rapidly spinning black holes \citep{blandford_active_1990, wilson_difference_1995}.
Therefore, mergers may be physically linked to radio loud AGN, and thus studies at those wavelengths will find a preference for merging host galaxies.
}

\textnormal{Observational studies of AGN overall conclude that there is no definitive universal merger-AGN connection, though the sample sizes in all studies are small ($\sim$ 10s - 100s of galaxies).
It appears to depend on the wavelength observed and redshift, as mergers are significant contributors to the AGN host population in IR and radio wavelengths and in local galaxies.
A complete comparison of the merger-AGN connection in TNG50 to observations requires modeling obscuration and IR emission from both galaxy and torus scale dust, a complex task outside the scope of this paper.
However, by applying Equation~\ref{eq:Lbol} for a simple comparison, we find little evolution of the fraction of AGN hosted by mergers with redshift when calculating the bolometric luminosity with  with major mergers hosting $\sim$30 percent of AGN with $L_{\textrm{bol}} > 10^{43}$ erg\,s$^{-1}$ across $0.2 < z < 3$ in TNG50. }

\FloatBarrier
\section{Conclusion} \label{Conclusion}
We have shown that galaxy mergers impact sSFR and sBHAR in Illustris TNG50 galaxies across the redshift range $0.2 \leq z \leq 3$.  
\textnormal{The high spatial and mass resolution of TNG50 allowed us to study the impact of mergers down to lower stellar masses than previous studies using the larger, lower resolution boxes. }
We additionally examined the effects of merger stage and merger mass ratio on sSFR and sBHAR. Our main conclusions are:
\begin{enumerate}
    \item \textnormal{We find stronger trends in excess sSFR or sBHAR with} galaxy stellar or black hole mass and redshift than with merger mass ratio or merger stage.
    \item At lower $z$ and $M_\star > 10^{9.5}M_\odot$, the difference between the merging and nonmerging populations is more pronounced, but mergers do not always have excess sSFR. Mergers are associated with modestly higher ($\sim1.5$x) sSFRs at both high-$z$ and $M_\star < 10^{9.5}M_\odot$. This trend holds for all merger stages and mass ratios (Figure ~\ref{fig:ssfrexcess}).
    \item Mergers are associated with modestly higher sBHARs ($\sim1.5-2.5$x) from $0.3 \lesssim z \lesssim 3$, \textnormal{except in the very lowest BH mass bin}.
    This trend holds for all merger stages and mass ratios (Figure~\ref{fig:sbharexcess}).
    \item The trend of elevated sSFRs and sBHARs in the merging sample becomes more pronounced overall as redshift decreases (Figures~\ref{fig:sSFRvsz} and \ref{fig:sBHARvsz}).
    \item Merging galaxies show elevated sSFRs and sBHARs for at least $\sim1$Gyr following the merging event (Figures~\ref{fig:TimeFromMergerSFR} and ~\ref{fig:TimeFromMergerBHAR}).
    \item Merging galaxies tend to have higher gas fractions than nonmerging galaxies throughout our entire redshift range (0.2 $\leq z \leq$ 3; Figure \ref{fig:gasfrac}). This higher gas content can provide the fuel for the elevated sSFRs and sBHARs.
\end{enumerate}

Cosmological simulations provide a tool to learn about galaxy mergers in regimes that are difficult to observe: high-$z$ and low stellar mass.
We have shown that in IllustrisTNG50, mergers do lead to enhanced star formation and black hole activity, and that factors such as stellar mass, merger mass ratio, and redshift all come into play in how much of an impact mergers have.
\textnormal{We will build on these results in an upcoming paper investigating the physical properties of galaxies that evolve with redshift to better understand the difference in how much mergers boost sSFR and sBHAR throughout time. Additionally, we will compare to common prescriptions for SFR and BHAR enhancements in mergers used in semi-analytic models.}
As \textit{JWST} and upcoming large, ground-based telescopes enable observations of early galaxies, we are filling in the picture of the role galaxy mergers play in cosmic star formation and AGN triggering. 
There have already been exciting discoveries of high-$z$  mergers \citep{gupta_mosel_2023} and AGN \citep{maiolino_jades_2023, greene_uncover_2024, lyu_active_2024, greene_uncover_2024} with \textit{JWST}.
Upcoming observations will provide further data to compare to cosmological simulations, thus providing further constraints \textnormal{on the modeling of physical processes such as star formation, black hole accretion, and black hole feedback.}

\begin{acknowledgements}
A.L.S. wishes to thank Robert Bickley, Jorge Moreno, and Kelsey Johnson for helpful discussions during the preparation of this work. 
A.L.S. and J.M.C. acknowledge support from NASA’s Astrophysics Data Analysis program, grant number 80NSSC21K0646, and NSF AST-1847938.
J.S. is supported by an NSF Astronomy and Astrophysics Postdoctoral Fellowship under award AST-2202388. The Flatiron Institute is supported by the Simons Foundation. 
\end{acknowledgements} 

\bibliography{references_zotero}{}
\bibliographystyle{aasjournal}

\end{document}